\documentclass[structabstract]{aa}  
\usepackage{graphicx}
\usepackage{txfonts,natbib}
\usepackage{color}
\begin{document}
   \title{Mixing at young ages: Beryllium abundances in cool main-sequence stars of the open clusters IC 2391 and IC 2602\thanks{Based 
on observations made with the ESO/VLT, at Paranal Observatory, under program 66.D-0284.}}


\author{R. Smiljanic
          \inst{1}
          \and
          S. Randich\inst{2}
          \and
          L. Pasquini\inst{1}
          }

   \institute{European Southern Observatory, Karl-Schwarzschild-Str. 2, 85748 Garching bei M\"unchen, Germany \\ 
                  \email{[rsmiljan,lpasquin]@eso.org}
   \and
   INAF - Osservatorio Astrofisico di Arcetri, Largo E. Fermi 5, 50125 Firenze, Italy \\
                   \email{randich@arcetri.astro.it}
             }

   \date{Received ; accepted }

 
  \abstract
   {The determination of lithium abundances in stars of young clusters have shown that they deplete Li by different degrees during their 
   pre-main sequence phase. Beryllium abundances are complementary to the lithium ones, and can help tracing the mixing processes 
   in the stellar interiors.}
   {Our aim is to derive beryllium abundances in a sample of G- and K-type stars of two young pre-main sequence open clusters, IC 2391 and 
   IC 2602. The Be abundances are used to investigate the mixing of internal material in these stars. The reliability of the Be lines as abundance 
   indicators in low-temperatures is also investigated in detail.
 }
   {We derived Be abundances from high-resolution, high signal-to-noise UVES/VLT spectra using spectrum synthesis 
   and model atmospheres. Atmospheric parameters and other elemental abundances are adopted from a previous work.}
   {The sample stars have masses in the range between 0.80 $\leq$ $M/M_{\odot}$ $\leq$ 1.20. They have been shown 
   to differ in lithium abundance by about 0.60 dex, with lower A(Li) in cooler and lower mass stars. Here, we find that all the stars 
   have the same Be abundance within the uncertainties. These observations show that the Be abundance is not affected 
   by the mixing events in the pre-main sequence, in this mass range, in agreement with the expectation of evolutionary models. 
   A comparison with Be abundances in older clusters shows that, contrary to the models, cool stars deplete Be during 
   their main-sequence lifetime, confirming what has been previously suggested in the literature.}
  {}
   

\authorrunning{Smiljanic, Randich \& Pasquini}

\titlerunning{Be abundances in IC 2391 and IC 2602}

   \maketitle
%

\section{Introduction}

Open clusters are ideal targets to study stellar evolution and stellar structure. They are composed 
of stars with the same age, chemical composition, and distance; moreover, their masses can be relatively well 
estimated. In young clusters, these advantages can be used to investigate the pre-main 
sequence (PMS) evolution of low-mass stars. 

In this context, the evolution of the light elements lithium and beryllium offer valuable information. 
These elements burn in (p,$\alpha$) reactions at relatively low but different temperatures 
in the stellar interiors ($\sim$ 2.5 $\times$ 10$^{6}$ K for Li and $\sim$ 3.5 $\times$ 10$^{6}$ K for Be). 
Variations in their surface abundances, with respect to their initial values, help constraining the physical 
processes mixing the surface material to that in the stellar interior \citep{Pin97,ipDPC00,ipCDP00}.

Li abundances have, consequently, been extensively studied in open clusters. 
According to standard evolutionary models\footnote{By standard evolutionary models we mean 
those that only allow for mixing in convective layers and do not include effects due to rotation 
and/or magnetic fields.}, Li depletion should be a unique function of age, mass, and chemical composition, 
in other words, within a given cluster stars with similar effective temperature should have undergone 
the same amount of Li depletion. Observations, however, have revealed a complex behavior with important discrepancies in almost all 
evolutionary stages of low- and intermediate-mass stars \citep[see e.g.][and references therein]{Pin97}. 

Focusing on PMS Li depletion, several surveys of young clusters have evidenced two major discrepancies 
between theoretical predictions and empirical patterns. First, standard models seem to predict more Li depletion 
than observed in PMS clusters (defined as clusters with age between $\sim$ 10 and 100 Myr). Second, there is 
a significant spread in Li abundances among late-G and K dwarfs, which has first been detected in the 120 Myr Pleiades 
\citep{DuncanJones83,Butler87}, along with a connection between Li abundances and rotation; namely, fast-rotating stars 
appear to have higher abundances than slow-rotating ones. This connection was confirmed in a large sample of 
Pleiades members by \citet{Soderblom93b} and has been found to hold for stars between 0.7 and 0.9 $M_{\odot}$, but to break down for 
less massive stars \citep{GarciaLopez94,Jones96}. The scatter in Li abundances in otherwise similar stars 
has since then been confirmed by other authors \citep{Jeffries99,Ford02,King10}.


Similar results of a star-to-star scatter in Li and the Li-rotation connection have been found 
in other clusters and associations both younger and older than the Pleiades, e.g. the nine young associations with ages 
between 6 and 70 Myr analyzed in \citet{DaSilva09}, IC 4665 \citep[$\sim$ 28 Myr,][]{MartinMontes97,Jeffries09}, 
IC 2602 \citep[$\sim$ 45 Myr,][]{Randich97,Ran01}, NGC 2451 A and B \citep[$\sim$ 50 Myr,][]{Hunsch04}, $\alpha$ Per 
\citep[$\sim$ 90 Myr,][]{Balachandran88,Balachandran11,RMLP98}, M 35 \citep[$\sim$ 175 Myr,][]{Barrado01}, NGC 6475 
\citep[$\sim$ 220 Myr,][]{JamesJeffries97,Sestito03}, and M 34 \citep[$\sim$ 250 Myr,][]{Jones97}. The general picture seems to indicate 
that the scatter decreases with age and disappears by the age of the Hyades 
\citep[$\sim$ 600 Myr,][]{Soderblom90,Soderblom95}. 

The interpretation of the Li scatter as a real abundance scatter and its connection to rotation is sometimes disputed in the literature 
\citep[see e.g.][]{King00,XiongDeng05}. It has been argued that the Li scatter is only an apparent effect caused by our failure of properly modeling 
the atmosphere of these stars, neglecting effects caused by stellar activity for example. Theoretical 
and observational investigations of these effects, however, seem to indicate that they are not able 
to fully explain the observed scatter \citep{Stuik97,Barrado01b,Randich01,King10}


We finally mention that the observed trend of Li with rotation is the opposite of that predicted when effects induced by rotation, 
such as meridional circulation and shear turbulence, are included \citep{Chaboyer95,Mendes99}. In the models, the fast-rotating 
stars are expected to deplete more Li then the slow-rotating ones. In the last 15 years several mechanisms related to 
rotation, magnetic fields, disk locking, and angular momentum loss have been proposed with the aim of 
solving the discrepancies with observed PMS Li depletion \citep{MartinClaret96,Ventura98,
Chabrier07,Bouvier08,BaraffeChabrier10,MacdonaldMullan10}. Whilst some of these models do predict a 
larger amount of Li depletion for slow rotators, they still need 
to be further investigated and constrained.

Beryllium, on the other hand, has been investigated in far fewer stars than Li. This is 
because Be abundances can only be determined from the \ion{Be}{ii} $^{2}$S--$^{2}$P$_{0}$ 
resonance lines at 3131.065 \AA\  and 3130.420 \AA, a spectral region near the atmospheric cutoff and 
thus strongly affected by atmospheric extinction. Moreover, this near-UV region is extremely crowded 
with atomic and molecular lines, some of them still lacking proper identification.

Most of the work on Be in open cluster has concentrated in F and early-G stars 
\citep[e.g.][]{SBKD97,BK02,BAK03b,BAK04,Ran02,Ran07,Sm10}, while very little has been done in 
cooler stars ($T_\mathrm{eff}$ $\leq$ 5400 K). Stars with temperatures down to $\sim$ 5000 K have 
been investigated only in the Hyades ($\sim$ 600 Myr) by \citet{GL95} and \citet{Ran07} and 
in NGC 2516 ($\sim$ 150 Myr) by \citet{Ran07}. 

These works have shown the following behavior of the Be abundances. As in the case of the Hyades,
 although the Li abundance decreases by two orders of magnitude with decreasing 
temperature between 6000 to 5400 K, the Be abundances remain unaltered \citep{BK02}. 
For cooler stars in the Hyades, the results of \citet{GL95} and \citet{Ran07} suggest a small 
degree of Be depletion.  For the four young stars in NGC 2516, no Be depletion was detected 
down to $\sim$ 5100 K \citep{Ran07}. For older field stars, \citet{San04b} have found some degree of 
Be depletion in stars cooler than 5600 K. These studies therefore suggest a low level of main-sequence 
Be depletion in the cooler stars.

Except for one star of IC 2391 analyzed by \citet{Ran02}, Be has never been systematically investigated 
in cool late G- and K-type stars of young pre-main sequence clusters. Here, we make for the first time such an 
attempt, and extend the investigation of Be to stars with $T_\mathrm{eff}$ as low as $\sim$ 4700 K in the 
PMS clusters IC 2391 and IC 2602. As evolutionary models fail to reproduce the Li depletion pattern in these 
clusters \citep{Ran01}, it is important to investigate Be in order to better constrain the properties 
of the mixing during the PMS phase. 

Our work is divided as follows. In Sec. \ref{sec:data} we 
describe the targets and the observational data. In Sec. \ref{sec:analysis} we present the model atmosphere and 
abundance analysis, discussing in detail the difficulties of modeling the spectra of rotating cool stars. Our 
findings are presented in comparison to evolutionary models and to literature data on older clusters in Sec. \ref{sec:diss}. 
Finally, in Sec. \ref{sec:final} we present our final remarks.

\section{Targets and observational data}\label{sec:data}

\subsection{The young clusters IC 2391 and IC 2602}

IC 2391 and IC 2602 are among the closest clusters to the Sun, with distances of the order of $\sim$150 pc 
\citep{vanLeeuwen09}. Although these two southern clusters are close to the Galactic 
plane, their proximity results in small reddening. The clusters also have very similar ages. For IC 2391, 
\citet{Barrado04} derived an age of 50 Myr using the lithium depletion boundary method \citep[see e.g.][]{Bildsten97}. 
For IC 2602, the Li depletion method has recently been applied by \citet{Dobbie10} and 
an age of 46 Myr was found. The clusters are young enough that 1 $M_{\odot}$ stars 
have just arrived or are arriving at the zero age main-sequence (ZAMS)\footnote{According to 
the models presented in \citet{PiauTurck02}, for example, stars with 1.4 $M_{\odot}$ and 1.0 $M_{\odot}$, 
with Pleiades chemical composition, arrive at the ZAMS after pre-main sequence lifetimes of $\sim$30 and $\sim$50 Myr, 
respectively.}. The data on the clusters is summarized in Tab. \ref{tab:opcl}. The young ages and 
proximity make the two clusters of considerable interest for the study of pre-main sequence evolutionary phases.  

\begin{table}
\caption{Physical data of the open clusters as adopted from the literature.} \label{tab:opcl}
\centering
\begin{tabular}{cccccc}
\noalign{\smallskip}
\hline\hline
\noalign{\smallskip}
Cluster & $(m-M)$ & $E(B-V)$ & Age & Distance & [Fe/H] \\
 & (mag.) & (mag.) & ( Myr) & (pc) &  \\
\hline
IC 2391 & 5.80 & 0.01 & 50 & 144.9 & $-$0.01 \\
IC 2602 & 5.86 & 0.04 & 46 & 148.6 &  0.00 \\
\noalign{\smallskip}
\hline
\end{tabular}
\tablefoot{The distance modulus and physical distances were determined using Hipparcos parallaxes by \citet{vanLeeuwen09}. 
The color excesses are from \citet{PattenSimon96} and \citet{Prosser96} for IC 2391 and IC 2602, respectively.  Their chemical 
composition was determined by \citet{DoraziRandich09}. The ages are discussed in the text.
}
\end{table}
%


In spite of that, only a few analyses of their chemical composition have been 
published. The metallicities were first determined 
by \citet{Ran01} for a subsample of the early-G and late-K type stars that they analyzed. 
Values of [Fe/H]\footnote{[A/B] = $\log$ [N(A)/N(B)]$_{\rm \star}$ $-$ $\log$
 [N(A)/N(B)]$_{\rm\odot}$} = $-$0.03 $\pm$ 0.07 and [Fe/H] = $-$0.05 $\pm$ 0.05 were found for 
IC 2391 and IC 2602, respectively. \citet{Stutz06} analyzed 5 B-F stars in IC 2391 including 
star SHJM2, one of the stars included in our sample. A mean metallicity above solar of 
[Fe/H] = +0.06 was determined, if we exclude HD 74044 which was found to be a slightly chemically 
peculiar A-type star. For SHJM2 a value of [Fe/H] = +0.08 was found. \citet{Platais07} derived an average value 
of [Fe/H] = +0.06 $\pm$ 0.06 from high resolution spectra of four slowly-rotating G-dwarfs. More recently, detailed abundances for 
15 G- and K-type dwarf members of the two clusters were obtained 
by \citet{DoraziRandich09}. Solar average metallicities of [Fe/H] = $-$0.01 $\pm$ 0.02 and [Fe/H] = 0.00 $\pm$ 0.01 
were determined for IC 2391 and IC 2602, respectively. Here we adopted these last values.

%
%

%
%

%
\begin{table*}
\caption{Observational data.} \label{tab:log} 
\centering
\begin{tabular}{lccccccc}
\noalign{\smallskip}
\hline\hline
\noalign{\smallskip}
Star (1) & Other & $V$  & $(B-V)$ & $(V-R)$ & Date of          & Exp. Time   &  S/N at  \\
          & Name & (mag.) (2) &  (mag.)  & (mag.)              & observation &  (s)                & 3130 \AA\ (3)    \\
\hline
VXR3a   & CD-52 2467                    & 10.95  &  \dots   & 0.37  & 16/02/2001 & 2  $\times$ 2400  & 100  \\
VXR31   & Cl* IC 2391 PMM 6478 & 11.22  &  \dots   & 0.37  & 17/02/2001 & 3 $\times$ 2400  & 20  \\
VXR67a & Cl* IC 2391 PMM 5859 & 11.71  &  \dots   & 0.52  & 16/02/2001 & 4 $\times$ 1800 &  55 \\
VXR70   & CD-52 2524                    &  10.85 &   \dots   & 0.38  & 17/02/2001 & 2  $\times$ 2400 &  115 \\
VXR72   & TYC 8569-230-1            & 11.46  &   \dots   & 0.43  & 18/02/2001 & 4 $\times$ 1800 &  70  \\
SHJM2   & CPD-52 1614                    & 10.30  & 0.57 & 0.33 & 16/02/2001 & 2  $\times$ 1200 & 90  \\
\hline
R1   & TYC 8960-1957-1 & 11.57 & 0.91 & \dots & 16/02/2001 & 4 $\times$ 1800 &  75  \\
R14 & TYC 8964-606-1   & 11.57 & 0.87 & \dots & 17-18/02/2001 & 4 $\times$ 1800 + 1 $\times$ 1650 &  75  \\
R15 & TYC 8964-73-1     & 11.75 & 0.93 & \dots & 18/02/2001 & 5 $\times$ 1800  &  60 \\
R66 & CPD-63 1624        & 11.07 & 0.68 & \dots & 16/02/2001 & 2 $\times$ 1800 &  80 \\
R70 & HD 307926            & 10.92 & 0.69 & \dots & 18/02/2001& 2 $\times$ 2400 &  80 \\
R92 & HD 308013            & 10.26 & 0.67 & \dots & 16/02/2001& 2 $\times$ 1200 &  70 \\
\noalign{\smallskip}
\hline
\end{tabular}
\tablefoot{
The stars from IC 2391 are listed in the upper part of the table and stars from IC 2602 in the lower part. (1) Identifiers for IC 2391 are from 
\citet[][VXR\#]{PattenSimon96} and from \citet[][SHJM\#]{Stauffer89}. For IC 2602 they are from \citet{Randich95}. 
(2) The VR photometry of the IC 2391 members is from \citet{PattenSimon96} and the (B$-$V) color of SHJM2 is from \citet{Stauffer89}. The BV 
photometry of the IC 2602 members is from \citet{Prosser96} and \citet{Randich95}. (3) Average S/N around 3130 \AA\ in the combined spectrum. 
}
\end{table*}
\begin{figure*}
\centering
\includegraphics[width=7cm]{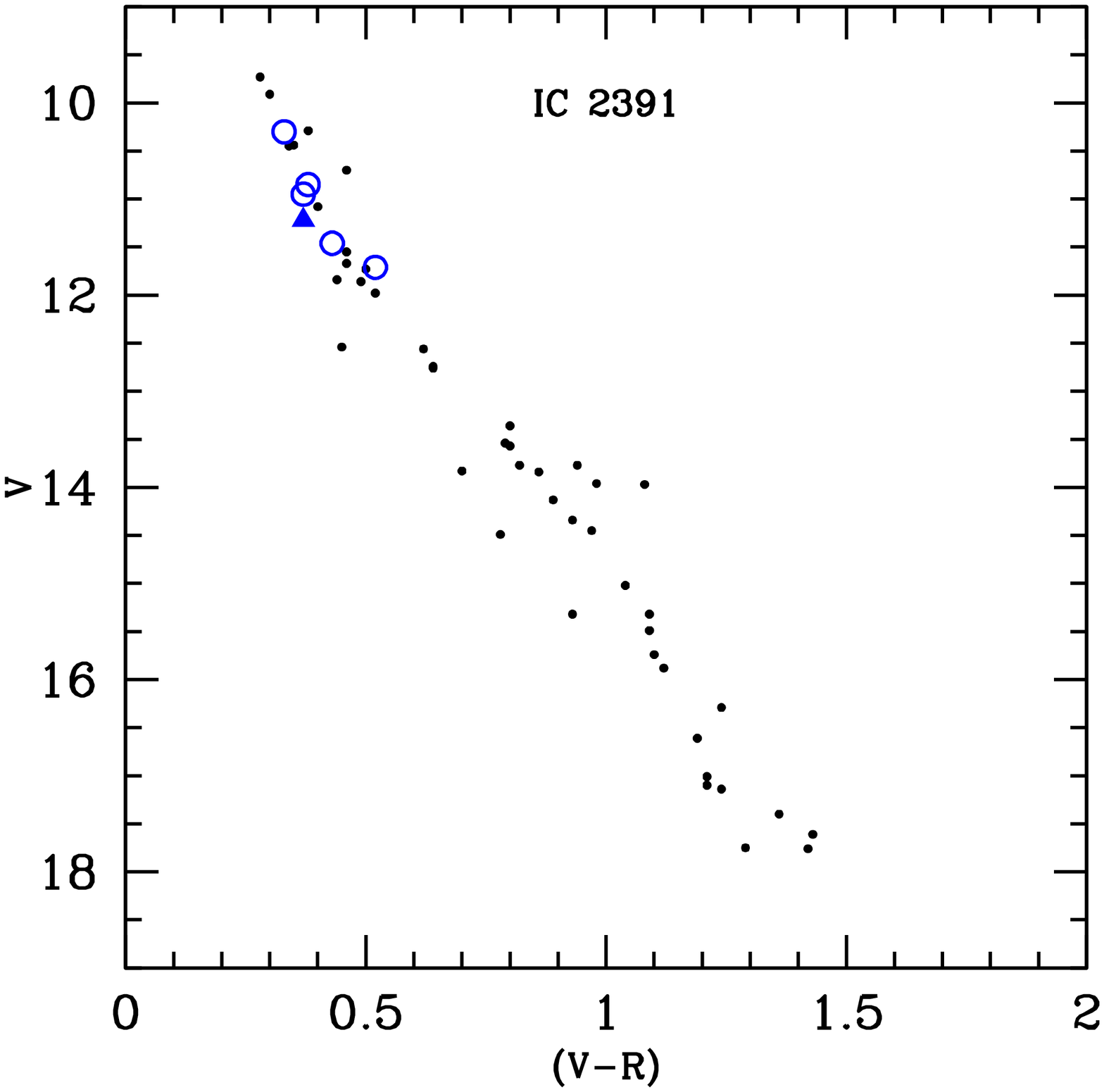}
\includegraphics[width=7cm]{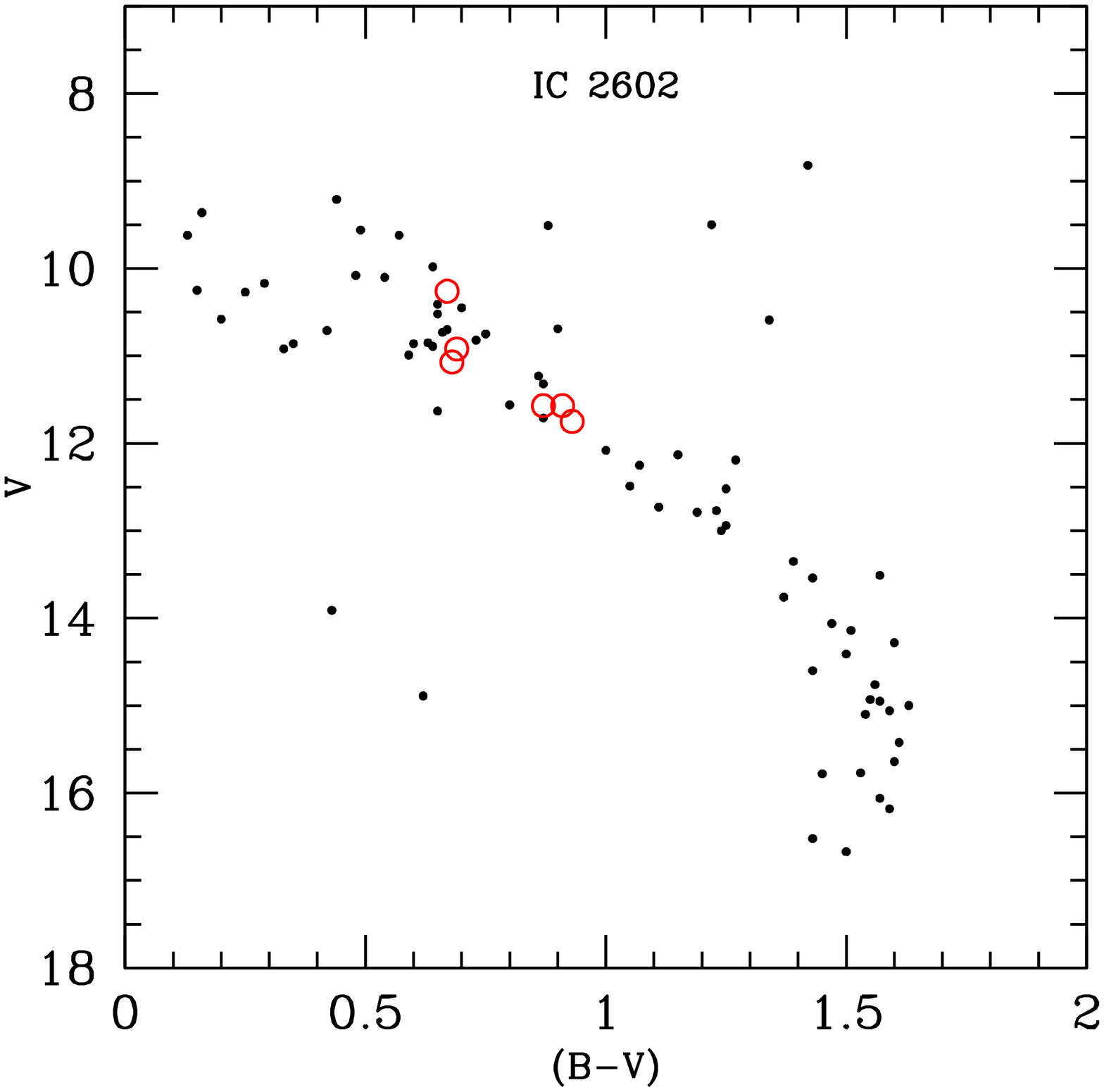}
\caption{Color magnitude diagrams of IC 2391 (left panel) and IC2601 (right panel). The observed stars 
are shown as open circles. The VR photometry of IC 2391 is from \citet{PattenSimon96}. Only members and 
suspected members according to the analysis of \citet{PattenSimon96} are shown. The BV photometry 
of IC 2602 is from \citet{Prosser96} and \citet{Randich95}. The photometry was obtained by means of the 
WEBDA database. In the CMD of IC 2391, star VXR31, found to be a non-member, is shown as a full triangle.}
\label{fig:cmds}
\end{figure*}

\subsection{Targets}

In this work we analyze six G- and K-type stars of each cluster (listed in Tab. \ref{tab:log}). All the stars were included in the 
work of \citet{DoraziRandich09}. The color-magnitude diagrams of the clusters are shown in Fig. \ref{fig:cmds}, 
where the observed stars are shown as open circles. The sample stars have been identified as photometric counterparts of possible 
low-mass cluster members detected in {\it ROSAT} X-ray observations \citep{Randich95,PattenSimon96}. The membership 
of these stars is supported by spectroscopic observations of radial velocities, H$\alpha$ activity, 
and Li abundances \citep{Randich97,Ran01,Stauffer97,Mermilliod09}. For IC 2391, proper motions 
have been measured by \citet{Platais07}. These measurements confirm the membership of all our sample stars, 
except for star VXR31, which is found to be a non-member. A search in the literature shows that this star 
has always been flagged as having a discrepant radial velocity, but was suspected to be a single-lined spectroscopic binary. 
We therefore assume here that it is not a member of the cluster, and we exclude it from further discussion. 




%
\begin{table*}
\caption{Atmospheric parameters and abundances of Li and Be. } \label{tab:par} \centering
\begin{tabular}{lcccccc}
\noalign{\smallskip}
\hline\hline
\noalign{\smallskip}
Star & $T_\mathrm{eff}$ (K) & $\log$ g   &  $\xi$  & $v~\sin i$& A(Li)  & A(Be) \\
    &   (K)   &    & (km s$^{-1}$) & (km s$^{-1}$) &  &  \\  
\hline
VXR3a   & 5590 & 4.45 & 1.15 & 10 & 2.80 & 1.30 \\
VXR67a & 4680 & 4.50 & 0.70 &  8  & 2.30 & 1.30 \\
VXR70   & 5557 & 4.50 & 1.15 & 17 & 2.80 & 1.30 \\
VXR72   & 5257 & 4.50 & 1.20 & 15 & 2.70 & 1.10 \\
SHJM2   & 5970 & 4.45 & 1.20 & $\leq$ 15 & 2.90 & 1.20 \\
\hline
R1   & 5050 & 4.50 & 1.15 & $\leq$ 10 & 2.40 & 1.25 \\
R14 & 5100 & 4.50 & 1.10 & 13 & 2.65 & 1.20 \\
R15 & 4770 & 4.50 & 0.70 & 7 & 2.40 & \dots \\
R66 & 5590 & 4.45 & 1.15 & 12 & 2.80 & 1.30 \\
R70 & 5760 & 4.45 & 1.10 & 12 & 2.90 & 1.15 \\
R92 & 5630 & 4.45 & 1.20 & 14 & 2.93 & 1.20 \\
\hline
vB109 & 5141 & 4.50 & 0.86 & \dots & 0.49 & 0.95 \\
vB182 & 5079 & 4.50 & 0.84 & \dots & 0.73 & 0.95 \\
\noalign{\smallskip}
\hline
\end{tabular}
\tablefoot{ 
The atmospheric parameters for the stars of IC 2391 and IC 2602 were derived by \citet{DoraziRandich09}. In the bottom we include two star from the 
Hyades with parameters determined by \citet{Ran07}. The rotational velocities are from \citet{Stauffer97}. The lithium abundances for the stars of 
IC 2391 and IC 2602 were calculated by \citet{Ran01} and include NLTE corrections. For the Hyades stars the Li abundances are from \citet{Ran07}. 
All the beryllium abundances were determined in this work. In addition, its spectrum has quite low S/N ratio and thus the value of A(Be) is quite uncertain.
}
\end{table*}
%

\subsection{Observational data}

The observations were carried out in visitor mode using UVES, the \emph{Ultraviolet 
and Visual Echelle Spectrograph} \citep{ipDe00} fed 
by UT2 of the VLT. UVES is a cross-dispersed echelle spectrograph 
able to obtain spectra from the atmospheric cut-off at 300 nm to 
$\sim$ 1100 nm. UVES was operated in dichroic mode with cross dispersers
\#1 and \#4 and central wavelengths 346 nm and 750 nm, respectively. A slit 
of 0.8$\arcsec$ was used, resulting in a resolution of R $\sim$ 55\,000 in the blue 
arm. The reduction was conducted with the ESO UVES data reduction 
pipeline within MIDAS \citep{uvespipeline}. The spectra have average signal-to-noise (S/N) between 
55 and 115 in the region around the Be lines. The log book of the observations 
is given in Table \ref{tab:log}.

%

\section{Analysis}\label{sec:analysis}

\subsection{Atmospheric parameters}

The atmospheric parameters were adopted from the work of \citet{DoraziRandich09}. 
Their analysis was done in LTE and differentially to the Sun. Effective temperatures 
were determined with the excitation equilibrium of \ion{Fe}{i} lines. A value of $\log$ $g$ = 4.50 
was in general adopted for the surface gravities, except for a few slow-rotating stars where the 
gravity could be checked using the ionization equilibrium between \ion{Fe}{i} and \ion{Fe}{ii} 
lines. In these cases a final value of $\log$ $g$ = 4.45 was found, in excellent agreement with 
the initial assumption. The microturbulence was constrained by requiring the abundances 
calculated from \ion{Fe}{i} lines to have a null correlation with the equivalent widths. The reader 
is referred to their work for more details on the adopted line list and atomic data. For the Sun, 
adopting the parameters, $T_\mathrm{eff}$ = 5777 K, $\log$ $g$ = 4.44, $\xi$ = 1.10 km s$^{-1}$, 
a value of A(Fe) = 7.52 was derived. As mentioned above, average metallicities of [Fe/H] = 
$-$0.01 $\pm$ 0.02 and [Fe/H] = 0.00 $\pm$ 0.01 were determined for IC 2391 and IC 2602, 
respectively. The atmospheric parameters are given in Tab. \ref{tab:par}. \citet{DoraziRandich09} 
estimate that the random errors affecting their parameters are of $\pm$ 60 K for $T_\mathrm{eff}$, $\pm$ 0.20 
km s$^{-1}$ for $\xi$, and between $\pm$ 0.15 and $\pm$ 0.25 dex for $\log$ $g$.

\subsection{Be abundances}

The beryllium abundances were determined using synthetic spectra calculated
with the codes described in \cite{CB05}. Model atmospheres were computed for the 
exact atmospheric parameters of the stars using the Linux version \citep{Sbordone04,Sbordone05} of 
the ATLAS9 code originally developed by Kurucz \citep[see e.g.][]{Kuruczcd13}. For the calculations we adopted 
the new opacities distribution functions of \citet{ipCK03} without overshooting. These
models assume local thermodynamic equilibrium, plane-parallel geometry, and hydrostatic equilibrium.

The line list is the same as used in our previous works on Be abundances \citep{Sm08,Sm09,Sm10}. 
The molecular line list is that described in \citet{CB05} and the atomic line list is the one compiled 
by \citet{Pr97}. Log $gf$ of $-$0.168 and $-$0.468 were adopted for the Be lines at 3131.066 \AA\ and 
3130.421 \AA, respectively. The line list includes a \ion{Fe}{i} line at 3131.043 \AA, with
$\log$ $gf$ = $-$2.517 and $\chi$ = 2.85 eV, with the objective of simulating an unknown 
line that affects the blue wing of the Be 3131 line. The parameters 
of this line were constrained using several stars of different parameters and metallicities \citep[see][for details]{Pr97}. 
The proper identification of this line is still controversial in the literature, and we remark that other choices have been 
made by different authors. As discussed in more detail by \citet{Ran07}, this choice does not affect the 
calculation of Be abundances of stars with $T_\mathrm{eff}$  $>$ 5400 K. As a number of our stars are cooler than that 
we discuss this issue in detail below.

With the line list described above, the solar UVES spectrum\footnote{The spectrum is available for download at the ESO website: www.
eso.org/observing/dfo/quality/UVES/pipeline/solar\_spectrum.html}, and adopting the parameters: $T_\mathrm{eff}$ = 5777 K, $\log$ $g$ = 4.44, 
and $\xi$ = 1.00 km s$^{-1}$ an abundance of A(Be)\footnote{A(Be) = $\log$ N(Be) = $\log$(Be/H) + 12.} = 1.10 is obtained. This abundance is in
excellent agreement with the one found by \citet{Chm75}, A(Be) = 1.15, and usually adopted as 
the reference photospheric solar abundance.

\begin{figure}
\begin{centering}
\includegraphics[width=7cm]{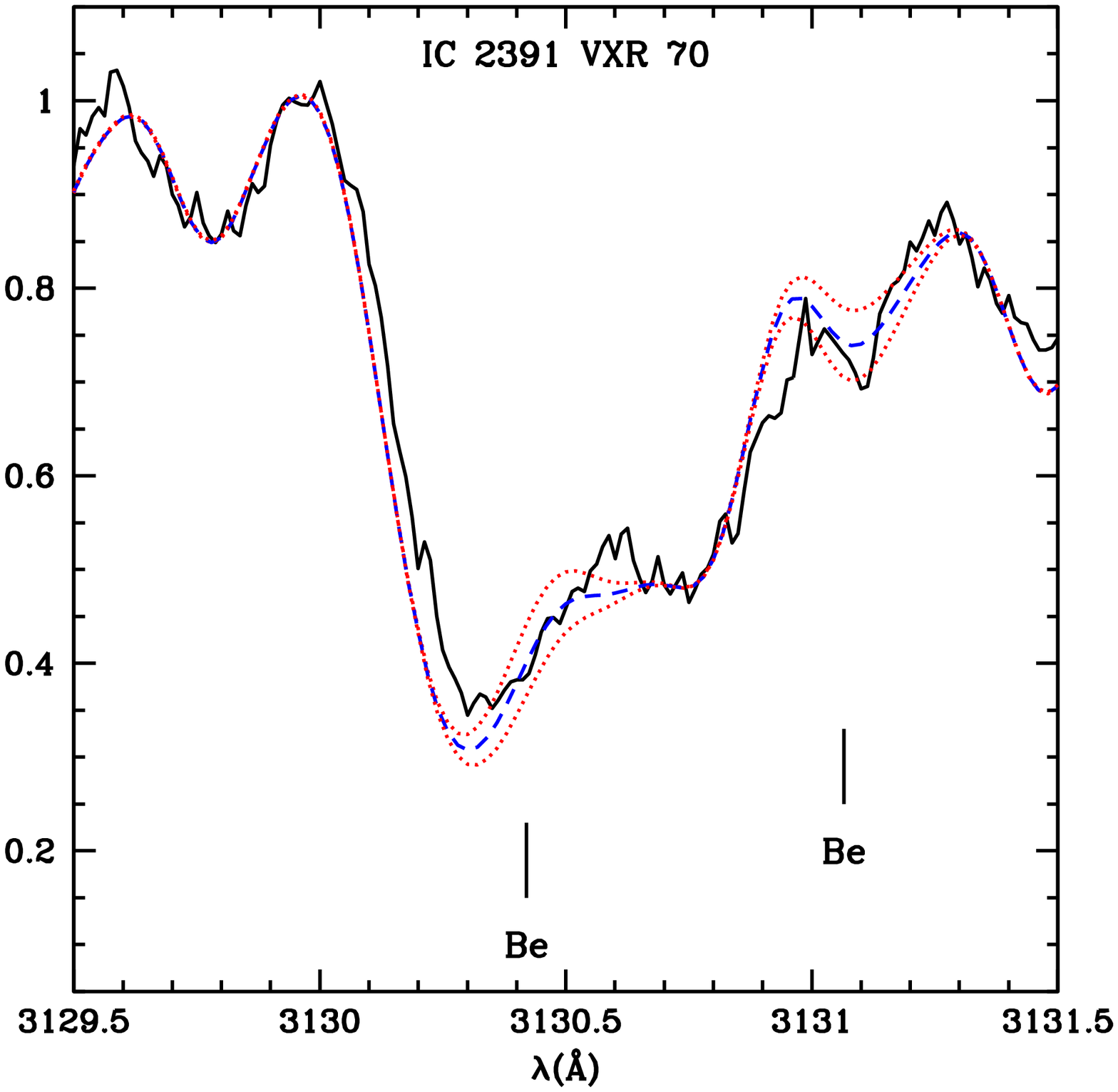}
\caption{Fit to the region of the Be lines in the star IC 2391 VXR 70. The solid line represents the observed spectrum, the 
dashed line a synthetic one with A(Be) = 1.30, and the dotted lines spectra with $\pm$ 0.20 dex in the Be abundance.} 
\label{fig:vxr70}
\end{centering}
\end{figure}

\begin{figure}
\begin{centering}
\includegraphics[width=7cm]{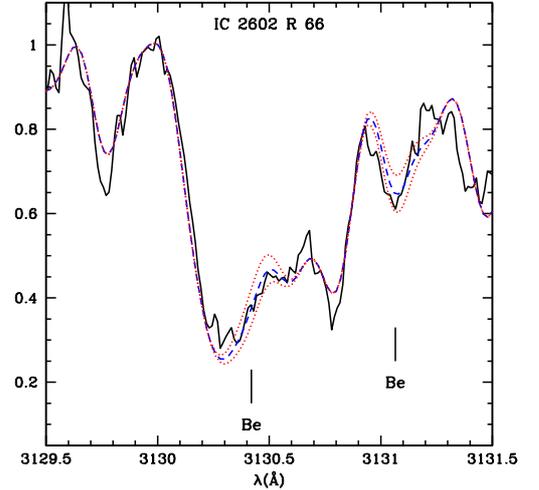}
\caption{Fit to the region of the Be lines in the star IC 2602 R 66. The solid line represents the observed spectrum, the 
dashed line a synthetic one with A(Be) = 1.30, and the dotted lines spectra with $\pm$ 0.20 dex in the Be abundance.} 
\label{fig:r66}
\end{centering}
\end{figure}

As for some stars in the open cluster IC 4651 analyzed in \citet{Sm10}, the spectra of 
our sample stars are affected by rotational broadening. \citet{Sm10} found that from this kind of 
data reliable, albeit slightly overestimated, Be abundances can be determined. The reader is 
referred to the discussion presented in \citet{Sm10} for more details. Examples of the fits for the stars 
VXR70 (from IC 2391) and R66 (from IC 2602) are shown in Figs. \ref{fig:vxr70} and \ref{fig:r66}, respectively.

We also derived Be abundances in two stars of the Hyades analyzed in \citet{Ran07}, vb109 and vb182. 
The abundance we find for both of them is A(Be) = 0.95 $\pm$ 0.23 (the two stars have temperatures of 5141 and 5079 K, respectively). That should be compared with 
A(Be) = 0.80 $\pm$ 0.21 derived in \citet{Ran07}. 




%

\subsection{The Be 3131 line below 5400 K}\label{sec:line}

Of the two Be resonance lines, the one at 3130 \AA\ is stronger but also more affected by blends. Because of that, the 
line at 3131 \AA\ is usually preferred for abundance determinations. The latter is, however, also affected by blends. 
The most important blends are: {\it (i)} a \ion{Zr}{i} line at 3131.109 \AA\ at its red wing \citep[included in 
our analysis with parameters determined by][]{CorlissBozman62}; {\it (ii)} two OH (A$^{2}\Sigma$-X$^{2}\Pi$) lines and 
one CH (C$^{2}\Sigma$-X$^{2}\Pi$) line \citep[included with parameters determined by][]{Cast99} at its blue wing; {\it (iii)} 
an unknown line at its blue wing that we model as a \ion{Fe}{i} line following \citet{Pr97}.

The lower the effective temperature the weaker the \ion{Be}{ii} lines are expected to become, as the population of the ionized species 
decreases. The blending features from neutral atoms and molecules, however, increase in importance. One might then question down 
to what temperatures the Be lines are still reliable abundance indicators.

To address this issue, the behavior of the 3131 Be line in late-type stars cooler than 5400 K was 
investigated in detail first by \citet{GL95} and then by \citet{Ran07}. \citet{GL95} adopted a different solution for the unknown 
blending line, i.e. increasing the $\log$ $gf$ value of a \ion{Mn}{i} line at 3131.037 \AA\ by 
1.50 dex\footnote{ We note that other tentative identifications for the blend have been advocated in the literature. For example, \citet{KDB97} decided 
to enhance the $\log$ $gf$ of the \ion{Mn}{ii} 3131.015 \AA\ line by +1.726 dex. However, as also pointed out in the paper, they failed to fit the spectrum of 
Procyon, since this choice predicts a feature that is too strong. In addition, from our exercise we conclude that a feature which increases in strength with decreasing temperature is more adequate. The ionized line of Mn would have the opposite behavior.}. This \ion{Mn}{i} line is also included in our line list, but with the original $\log$ $gf$ determined by \citet{inpKurucz88}, 
i.e. $\log$ $gf$ = $-$1.725. \citet{GL95} concluded that below $\sim$ 5200 K, the \ion{Mn}{i} becomes the dominant component 
of the blend, therefore the utility of the 3131 Be line as a good abundance indicator decreases. The exact temperature where it might 
become useless, however, would depend on both the resolution and the S/N ratio of the spectrum. Because of this, \citet{GL95} 
were able to put only upper-limits to the Be abundance in their sample stars with $T_\mathrm{eff}$ $<$ 5200 K.

%

In \citet{Ran07}, a similar exercise was conducted. These authors, however, adopted the same atomic line list that we use in 
this work and compared the different effects caused by adopting either the \ion{Fe}{i} or the \ion{Mn}{i} line. \citet{Ran07} 
concluded that for stars above 5400 K choosing either of the solutions does not affect in a significant way the Be abundances. 
For lower temperatures, they found that the Be abundances can differ by 0.15$-$0.10 dex between 5250$-$5400 K, and up to 
0.20 dex at $\sim$ 5000 K. Smaller abundances are derived when the \ion{Fe}{i} line is adopted (i.e., this line is found to be 
always stronger than the Mn one). Using the Fe blend, \citet{Ran07} concluded that it was possible to measure Be down to 
the lowest $T_\mathrm{eff}$ in their sample, $\sim$ 5000 K. 

\begin{figure*}
\centering
\includegraphics[width=6cm]{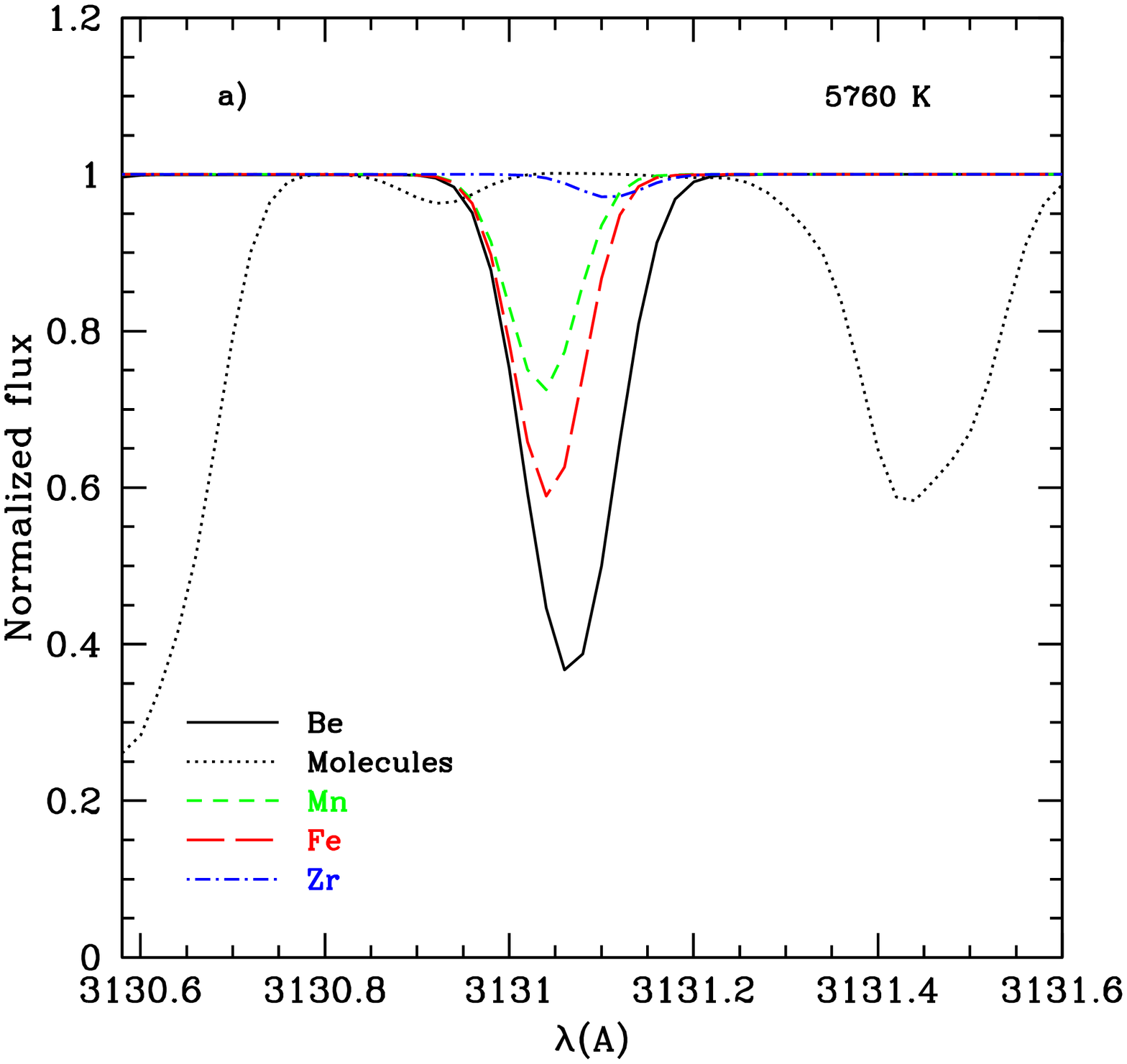}
\includegraphics[width=6cm]{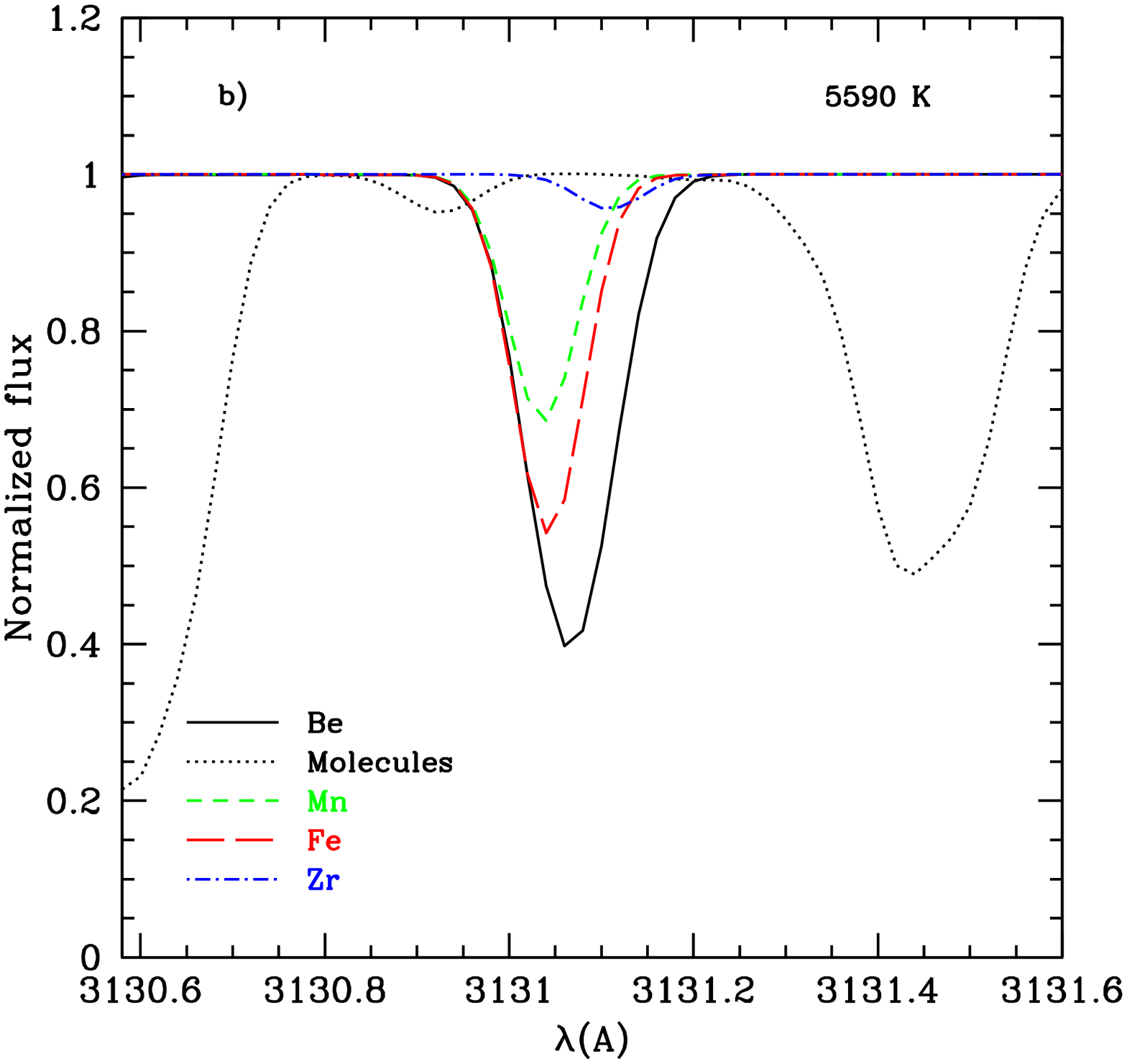}
\includegraphics[width=6cm]{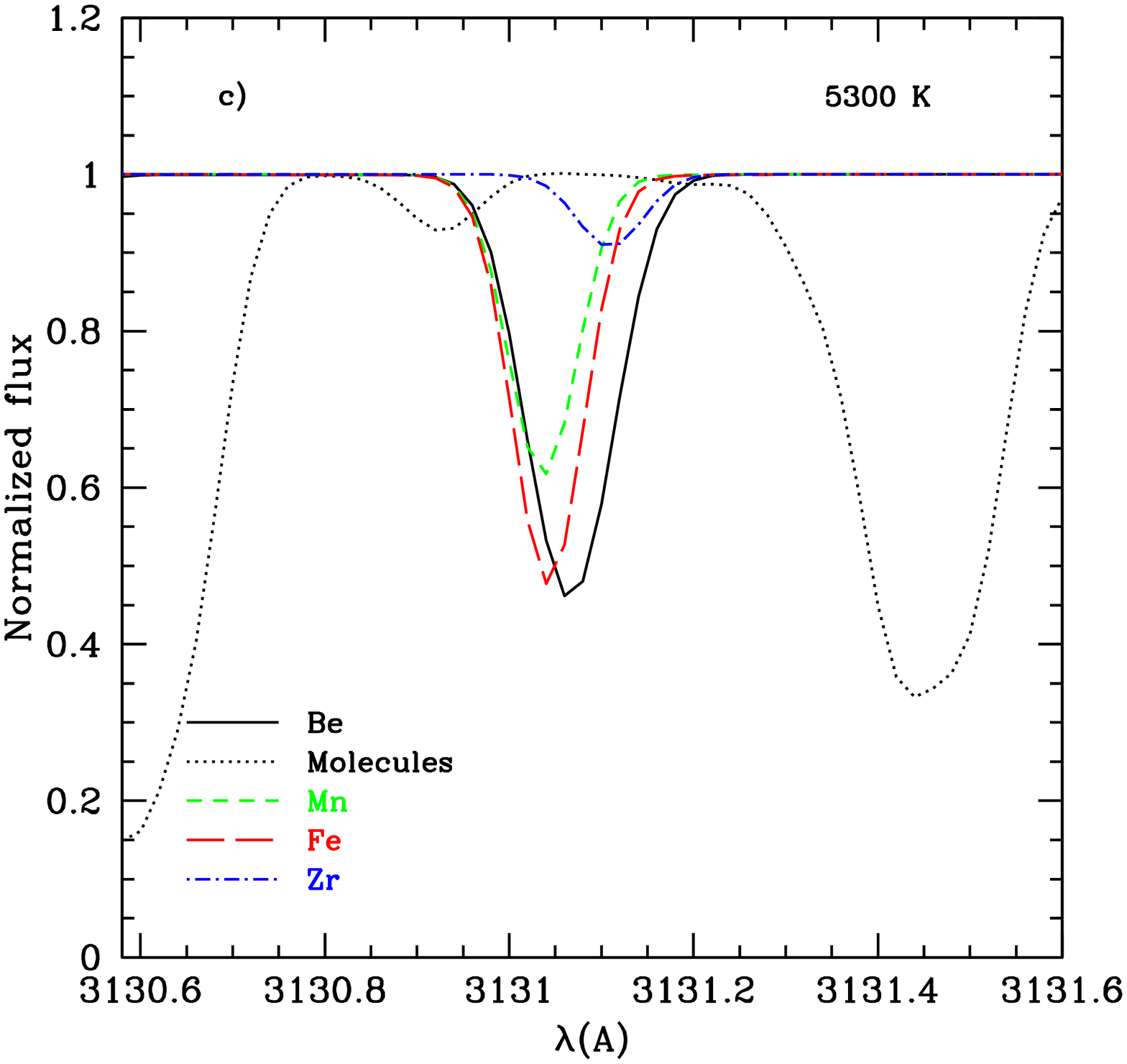}
\includegraphics[width=6cm]{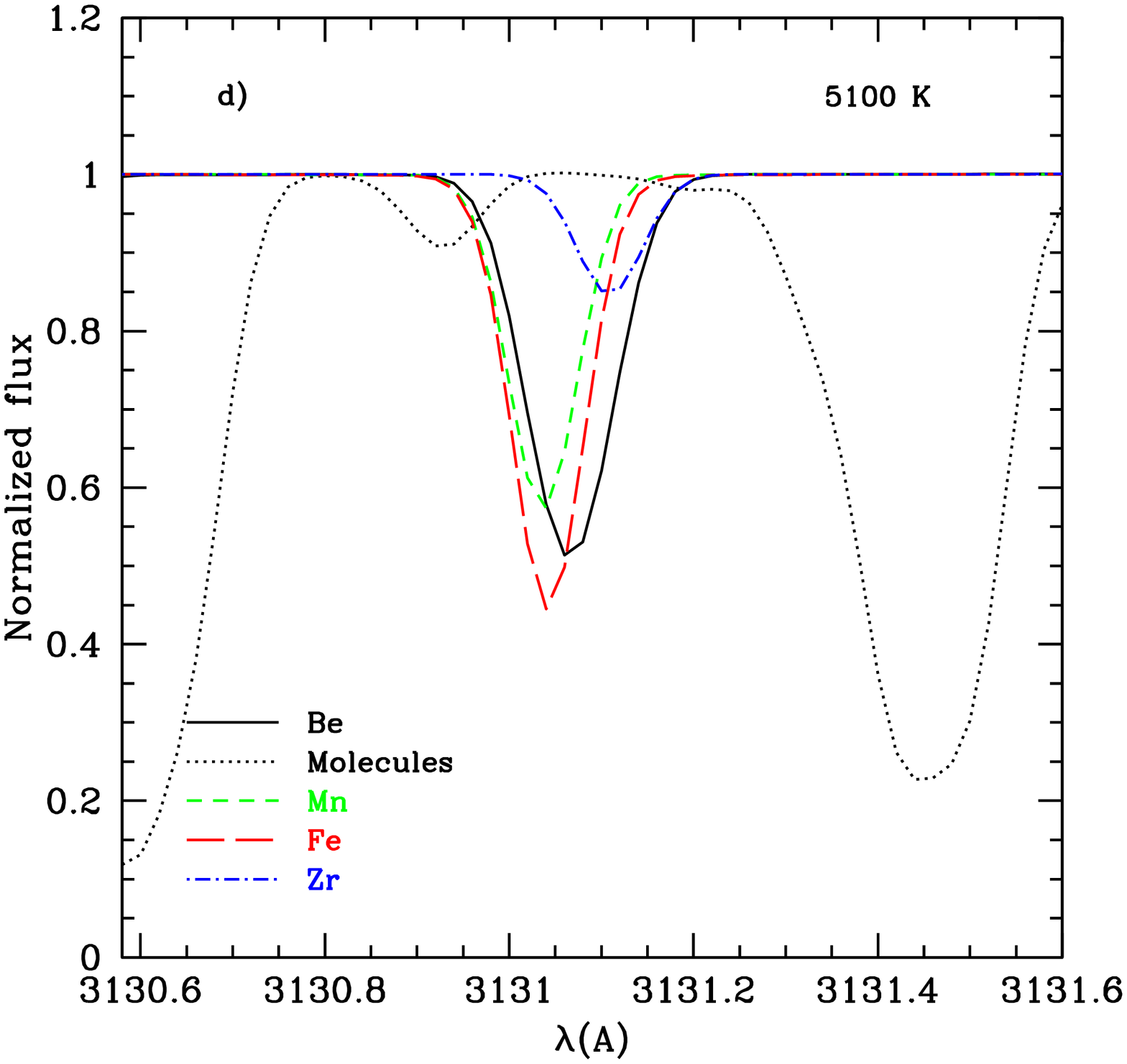}
\includegraphics[width=6cm]{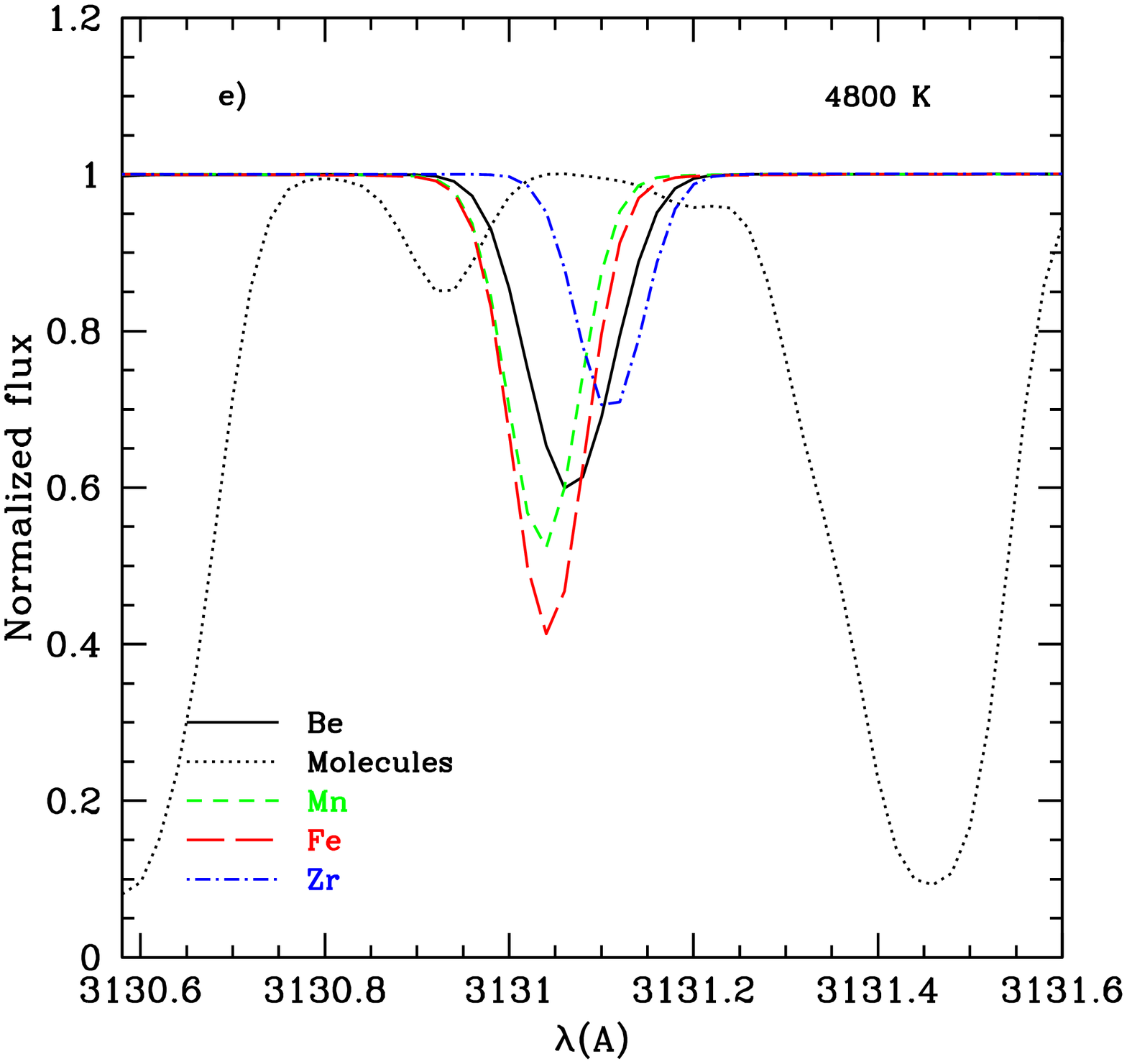}
\caption{Comparison of the strength of the lines affecting the Be line at 3131 \AA. The Be line is shown as a solid 
black line, the molecular features as a dotted black line, the Mn line (with increased $\log$ $gf$) as a green short-dashed line, the Fe line as 
a red long-dashed line, and the Zr line as a blue dot-dashed line. Five different temperatures, corresponding to the 
temperature range of our sample, are shown: (a) 5760 K, (b) 5590 K, (c) 5300 K, (d) 5100 K, and (e) 4800 K.}
\label{fig:lines}
\end{figure*}

Here, we repeat the exercise, comparing the Fe and Mn lines, but also investigating the influence of the \ion{Zr}{i} and 
the molecular lines on the blend and extending the test down to 4800 K. A comparison of the relative strengths of these 
lines is shown for five different temperatures in Fig. \ref{fig:lines}.

%

%
%

As expected, the \ion{Be}{ii} line becomes weaker with decreasing temperature while the lines 
of \ion{Fe}{i}, \ion{Mn}{i}, and \ion{Zr}{i} all become stronger. At $\sim$ 5100 K, the added contribution 
of the blending lines are comparable to the strength of the Be line, irrespective of the choice between Fe or Mn. 
As was found by \citet{Ran07}, the Fe line is always stronger than the Mn line. We also confirm the 
differences in the Be abundances caused by adopting different lines in the temperatures range studied by 
\citet{Ran07}. In addition, we find that for even lower temperatures, down to $\sim$ 4800 K, 
the difference can amount to 0.40$-$0.45 dex.

\begin{figure}
\begin{centering}
\includegraphics[width=7cm]{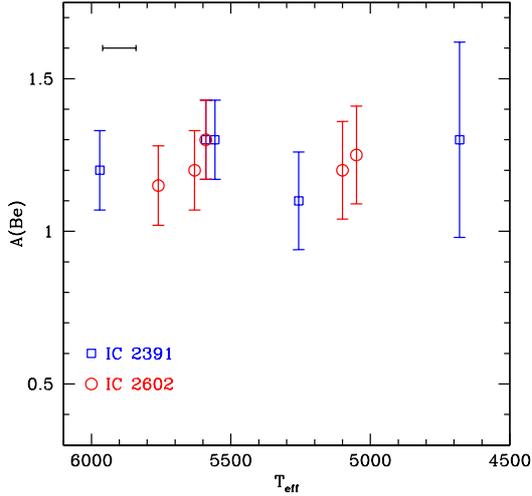}
\caption{Beryllium abundances as a function of the effective temperature of the stars in IC 2391 (open blue squares) and 
IC 2602 (open red circles). The error bar in temperature is shown only on the upper left side of the figure.}
\label{fig:teffbe}
\end{centering}
\end{figure}

While we have no definitive answer to whether the blend is caused by a Fe or a Mn line, it is clear that 
had we chosen to use the Mn line in our analysis, we would obtain a trend of increasing Be abundance with 
decreasing temperature (in Fig. \ref{fig:teffbe}), as we would need to compensate for the weaker blend by increasing 
the Be abundance to fit the 3131 line. This increase in the Be abundance would also make the 3130 line much stronger than observed. 
Choosing the Fe blending line, we can get a good simultaneous fits of both the 3130 and 3131 lines. Therefore, we conclude here that the Fe line seems 
to be a better solution for the blend, although in the future it might still be proven that it is not the right one.

\begin{figure}
\begin{centering}
\includegraphics[width=7cm]{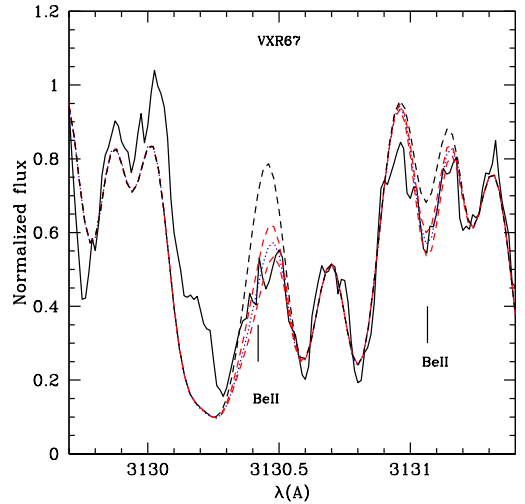}
\caption{Fit to the region of the Be lines in the star IC 2391 VXR 67 with $T_\mathrm{eff}$ = 4680 K. The solid line represents the observed spectrum, the 
dotted line a the best fit synthetic spectrum with A(Be) = 1.30, and the dashed lines represent spectra without Be and with $\pm$ 0.25 dex from the best fit.} 
\label{fig:vxr67}
\end{centering}
\end{figure}

Although the Be line is not the dominant line at low temperatures, the blend should still be sensitive to the Be abundance, 
as long as the blending lines are well constrained. Although this is not perfectly true, from the exercise we conducted 
we believe to have as good a control over the blends as is currently possible. Thus, using spectrum synthesis, 
Be abundances can be determined down to at least $T_\mathrm{eff}$ $\sim$ 4700 K, which is the low-temperature end of our sample. 
This is illustrated in Fig. \ref{fig:vxr67}, where we show the spectra of our coolest sample star in comparison with 
synthetic spectra with different Be abundances.

To finalize, two points must be stressed. First, for as long as the blends remain not well known, we can not claim to 
determine absolute Be abundances at the low $T_\mathrm{eff}$ range with sufficient precision. The relative comparison between similar stars in the same 
$T_\mathrm{eff}$ range, however, should be robust. Second, as the Fe line we adopt is the stronger among the possible 
choices, our Be abundances should be on the low-end. Therefore, our conclusion that there is no significant Be depletion in the stars analyzed here 
(see Section \ref{sec:diss} below) is robust.

\subsection{Uncertainties in the Be abundances}

The Be abundances are affected by random uncertainties coming from the atmospheric parameters and 
by uncertainties in the determination of the pseudo-continuum during the spectrum synthesis. To estimate 
the effect of the atmospheric parameters, we can change each one by its own error, keeping the other 
ones with the original adopted values, and recalculate the abundances. Thus, we measure how the variation 
of one parameter affects the abundances. We assume the uncertainties introduced 
by each parameter to be independent of the other ones. The uncertainty due to the continuum was determined by 
estimating the sensitivity of the derived Be abundance on the adopted continuum level. This uncertainty is 
mostly related to the S/N of the spectrum, and thus can be slightly different from star to star. Estimations of these effects 
are listed in Table \ref{tab:sigma}. The calculations were done for three stars, spanning the rage in temperatures of the 
sample (VXR67, R14, and SHJM2). In the plots that follow these uncertainties were considered as typical values also for 
the other sample stars with similar temperatures.

As mentioned in Sec. \ref{sec:data}, \citet{Ran02} has also analyzed one of  sample stars, star SHJM2 
of IC 2391. The atmospheric parameters they used are the same adopted here. An abundance of 
A(Be) = 1.11 $\pm$ 0.13 was derived. Our own value, A(Be) = 1.20 $\pm$ 0.13 agrees with theirs within 1$\sigma$. 

\begin{table}
\caption{Uncertainty of the Be abundance.}
\centering 
\label{tab:sigma}
\begin{tabular}{ccccccc}
\noalign{\smallskip}
\hline\hline
\noalign{\smallskip}
Star & $\sigma_{\rm Teff}$ & $\sigma_{\rm log g}$ & $\sigma_{\rm \xi}$ &
 $\sigma_{\rm [Fe/H]}$ & $\sigma_{\rm fitt.}$  & $\sigma_{\rm total}$ \\
\hline
SHJM2 & $\pm$ 0.05 & $\pm$ 0.10 & $\pm$ 0.00 & $\pm$ 0.03 & $\pm$ 0.05 &  $\pm$ 0.13 \\
R14       & $\pm$ 0.05 & $\pm$ 0.10 & $\pm$ 0.05 & $\pm$ 0.00 & $\pm$ 0.10 &  $\pm$ 0.16 \\ 
VXR67 & $\pm$ 0.10 & $\pm$ 0.20 & $\pm$ 0.10 & $\pm$ 0.00 & $\pm$ 0.20 & $\pm$ 0.32 \\
\noalign{\smallskip}
\hline
\end{tabular}
\tablefoot{ The effect introduced by the uncertainties of each of the atmospheric parameters and by the uncertainty of the fitting 
itself related to the level of the continuum and the S/N.
}
\end{table}

\subsection{Lithium abundances}

Lithium abundances for IC 2391 and IC 2602 were first calculated by \citet{Stauffer89} and 
\citet{Randich97}, respectively. The samples of member stars of these two clusters were then 
reanalyzed and extended by \citet{Ran01}. The Li abundances listed in Tab. \ref{tab:par} were 
all taken from this last reference. Li abundances are particularly sensitive to temperature and 
the $T_\mathrm{eff}$ adopted in \citet{Ran01} are for a few stars slightly different 
from the ones adopted here. The differences amount to at most 70 K. According to the analysis 
presented in \citet{Ran01}, these differences could result in 0.10$-$0.15 dex differences in the 
Li abundances. Differences that are within the uncertainties of the analysis itself (for our sample 
stars the uncertainties in A(Li) listed in \citet{Ran01} vary between 0.18 and 0.30 dex).

The Be abundances, on the other hand, are only weakly sensitive to the $T_\mathrm{eff}$. Because of 
this, and because the Li abundances from \citet{Ran01} are, anyway, consistent within themselves, 
we decided to just adopt their values for A(Li) instead of recomputing the abundances.

%

\section{Discussion}\label{sec:diss}

\subsection{Mixing in the pre-main sequence}

\begin{figure}
\begin{centering}
\includegraphics[width=7cm]{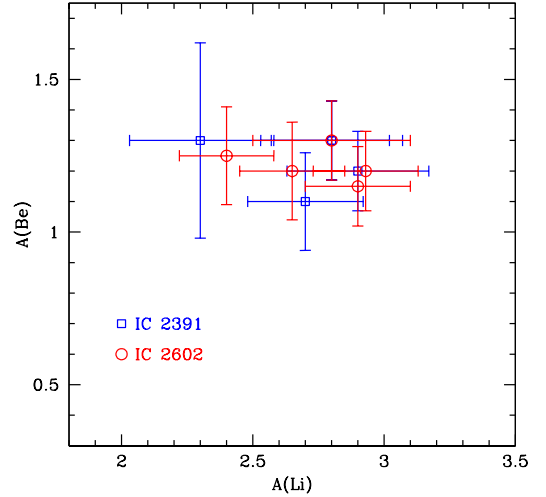}
\caption{Beryllium abundances as a function of the Li abundances derived by \citet{Ran01} for the stars in IC 2391 (open squares) and 
IC 2602 (open circles).}
\label{fig:beli}
\end{centering}
\end{figure}

Solar-type stars start their lifes as fully convective objects with large radii and low 
central temperature. As they contract, the stellar temperature rises and may eventually 
reach values high enough to burn the light elements. The rise in temperature can also 
lead to the development of a radiative core while the convective layer retreats to the 
upper layers. The radiative core will appear at later ages the smaller the mass and 
the higher the metallicity of the star. Light element depletion will only be observed if the 
temperature in the convective zone reaches the values necessary for nuclear burning. 
If the temperature is high enough for depletion only in the radiative core, no depletion 
is observed in the photosphere.


%
\begin{figure*}
\begin{centering}
\includegraphics[width=7cm]{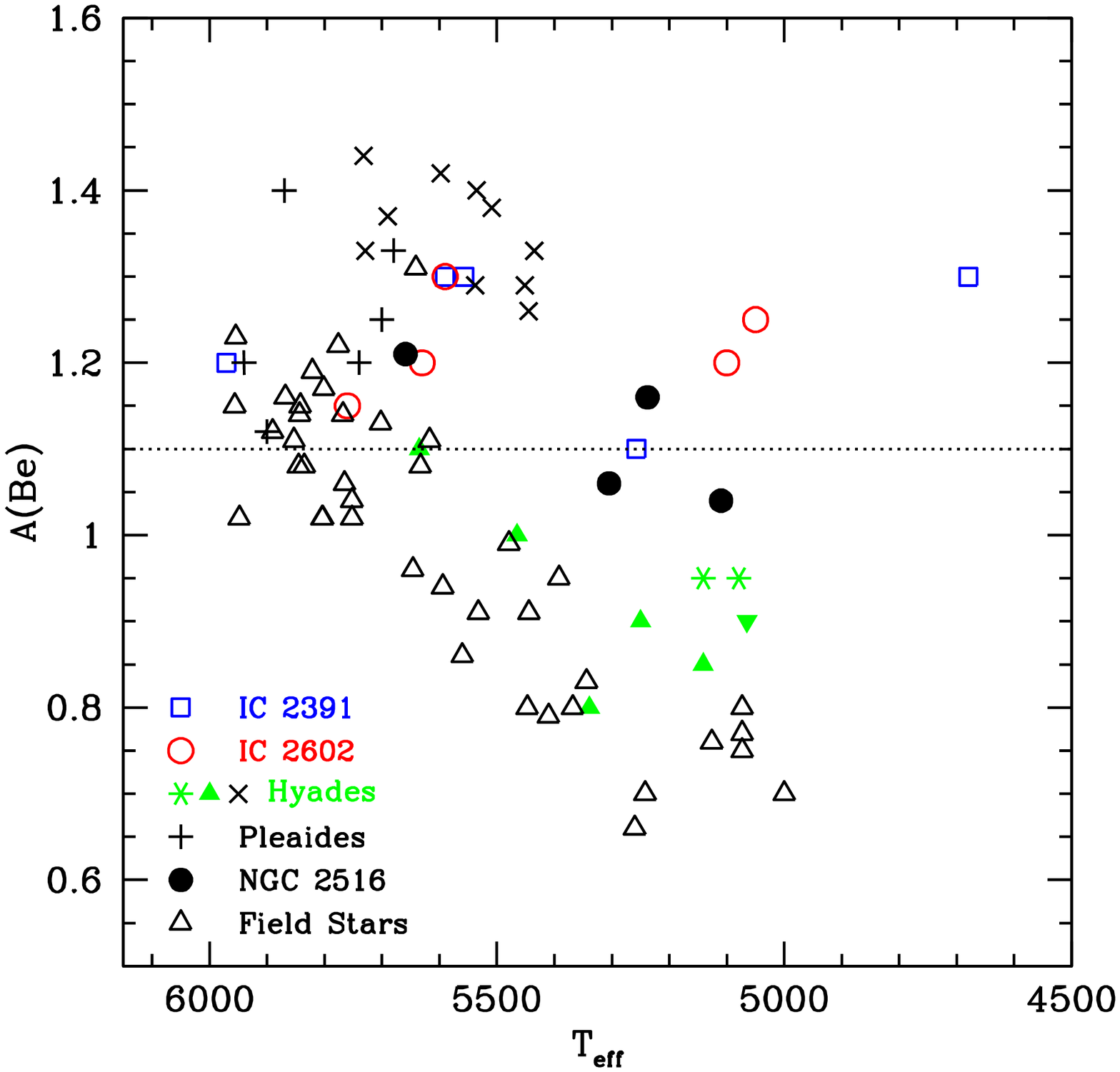}
\includegraphics[width=7cm]{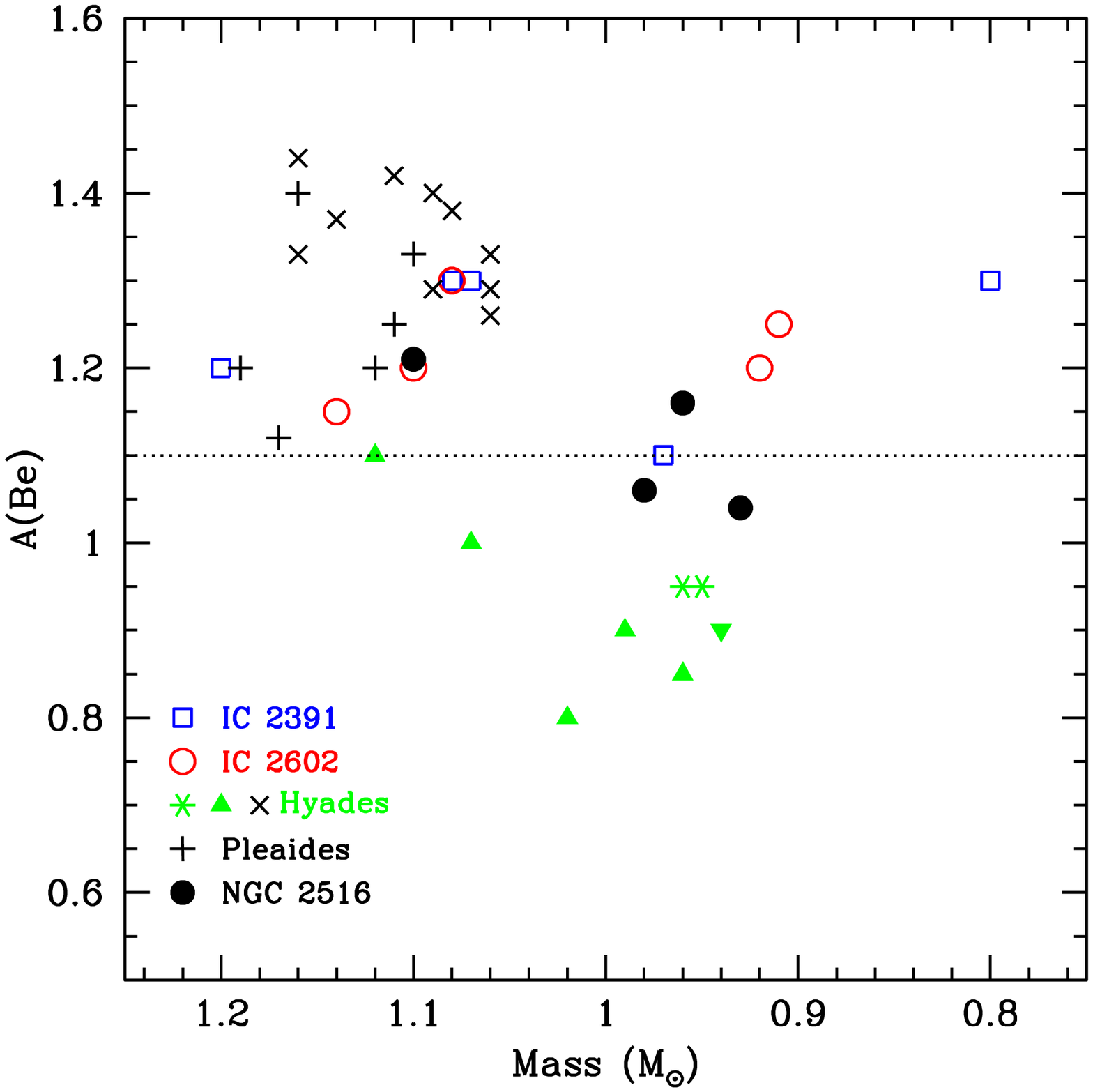}
\caption{Comparison with the Be abundances from selected results of the literature. In the left panel the abundances are shown as a function of $T_\mathrm{eff}$ and in the 
right panel as a function of mass. As in the previous figures, our results for IC 2391, IC 2602, and the Hyades are 
shown as open blue squares, open red circles, and green starred symbols, respectively. Abundances for additional Hyades stars collected from 
\citet{Ran07} and \citet{GL95} are shown 
as full green triangles. Abundances for Hyades stars from \citet{BK02} and from Pleiades stars from \citet{BAK03b} are shown as ``$\times$'' and ``$+$'' symbols, 
respectively. Abundances for NGC 2516 stars from \citet{Ran07} are shown as full black circles and abundances for field stars from \citet{San04b} are shown as open 
triangles. The dashed line indicates the solar value.}
\label{fig:lit}
\end{centering}
\end{figure*}

Models and observations have shown that stars more massive then $\sim$ 1.20 $M_{\odot}$ do 
not deplete Li during the PMS. The amount of depletion increases for less massive stars. At about 
0.6 $M_{\odot}$ all Li will be burned before the radiative core is able to develop. As beryllium is less 
fragile than Li, it is expected to survive longer. Significant Be depletion is expected by the models 
only for masses below $\sim$ 0.6 $M_{\odot}$ \citep{Bodenheimer66,ChabrierBaraffe97}. This expectation, however, has not been empirically 
tested before and this is our aim here.


Before discussing the Be abundances, lets briefly recall the behavior of the Li abundances 
in our two clusters. As discussed in \citet{Ran01}, stars more massive then $\sim$ 1.0 $M_{\odot}$ do not show signs of 
Li depletion, while cooler stars are depleted by an amount that increases with decreasing temperature. 
\citet{Ran01} concluded that the late-G and early-K stars in IC 2602 present a scatter in Li 
abundances that is similar, but not as large, as the one seen in the Pleiades. IC 2391 did 
not show strong evidence of a scatter, but this could be caused by low number statistics. 
Some correlation with rotation is seen, but not as strong as in the Pleiades.

The dependence of Li depletion on temperature was found to be similar in the two clusters 
down to $\sim$ 5000 K. At lower temperatures, the stars in IC 2391 seem to be more depleted 
than the ones in IC 2602. Moreover, some of the stars in the two clusters are more depleted in 
Li than similar stars in the older Pleiades.  These observations point that the pattern 
of depletion may be affected by the individual properties of each cluster.

The observed correlation between Li and rotation in main-sequence stars in young clusters is 
a strong indication that the evolution of stellar angular momentum during the PMS affects the 
efficiency with which Li is burned. In addition to the increase in rotation due to the contraction of the 
star, the angular momentum evolution is also affected by the interaction of magnetic fields 
with the disk and/or the stellar wind.

Given the complexity of the processes involved, abundances of both Be and Li are useful to probe the range in temperatures 
achieved in the convective zone during the PMS. Thus, demonstrating whether the expectation 
from the models, i.e. that Be is not affected, is important. 


In Fig. \ref{fig:teffbe} we show the Be abundances of the stars as 
function of the effective temperatures. The mean abundance of our stars is A(Be) = 
1.23 $\pm$ 0.07. No trend with temperature is detected. Taking into account that the 
spectra of many of our sample stars are affected by rotational broadening, the Be 
abundances we derive are expected to be slightly overestimated \citep[see 
discussion in][]{Sm10}. Therefore, given the uncertainties, there is no significant difference between the abundances measured 
in our stars and that observed in the Sun (where A(Be) = 1.10).

Recently, \citet{Takeda11} proposed that most early-G stars suffer some degree of gradual Be depletion. This conclusion is based on a 
weak correlation found between A(Be) and $v~\sin i$ in a large sample of solar analogues. However, the observations of Be in 
old open clusters and the similarity of the Be abundances among cluster members of different ages and field 
stars seem to argue against that, suggesting instead that no Be depletion has taken 
place in stars between 6100 K $\leq$ $T_\mathrm{eff}$ $\leq$ 5600 K. If that is indeed the 
case, then Fig. \ref{fig:teffbe} strongly suggests that our cooler stars also did not suffer 
any Be depletion. Below 5000 K, we were only able to determine the Be abundance 
in one star, VXR67 (see Fig. \ref{fig:vxr67}). Although its abundance has a larger error bar, 
it is fully consistent with the remaining objects. There is no indication of Be depletion.


This conclusion is reinforced by Fig. \ref{fig:beli}, where we plot the Be abundances 
as a function of the Li abundances derived in \citet{Ran01}. In the case where Be is 
affected by PMS depletion, one would expect the abundances of Li and Be to 
correlate, i.e. that stronger Be depletion should be seen in stars with stronger Li 
depletion. As no such trend is seen in Fig. \ref{fig:beli} there is again no indication 
of Be depletion.

We recall here that, as discussed in Sec. \ref{sec:line}, the Fe line we adopt as a blend is the stronger 
among the possible choices. Therefore, any systematic effect introduced by this choice would result 
in a smaller Be abundance for cooler stars. Thus our conclusion that there is no significant Be depletion 
in the stars analyzed here is robust.




\subsection{Depletion on the main sequence}

In Fig. \ref{fig:lit} we present a comparison of our results with Be abundances of other clusters and field stars. The references from where 
the abundances were compiled are given in the caption of the figure. All of them derived a solar abundance of beryllium similar to the one 
we obtain here, and thus are in a similar scale. Exception are the results of Boesgaard and collaborators who found A(Be)$_{\odot}$ = 
1.30 \citep{BAK03b}. If we were to express the results using the Sun as reference, we would be in the same scale. The addition of their 
results in Fig. \ref{fig:lit} do change our conclusion below, as their stars are located in the same temperature range where all other 
results also do not find Be depletion. 

The masses of the stars shown in the left panel of Fig. \ref{fig:lit} were interpolated from theoretical isochrones, for the age of 
the cluster and the given temperature of the star. The isochrones were calculated with the internet server\footnote{http://www.astro.ulb.ac.be/{$\sim$}siess/server/iso.html} dedicated to pre-main sequence and main-sequence model calculations presented by \citet{Siess00}. 
From these isochrones, it seems that the masses of stars older than the ZAMS might 
be slightly overestimated (by $\sim$ 0.10 $M_{\odot}$).

The sequence from the clusters to the field in Fig. \ref{fig:lit}  can be seen as a sequence of age: from IC 2391 and IC 2602 ($\sim$ 50 Myr), to 
NGC 2516 ($\sim$ 150 Myr), the Hyades ($\sim$ 600 Myr), to 
the field stars of \citet{San04b} (older than $\sim$ 1 Gyr). It is seen that for stars cooler than $\sim$ 5600 K 
(mass lower than $\sim$ 1.10 $M_{\odot}$) the degree of Be depletion seems to increase with age. 
There is also a clear correlation between Be depletion and temperature (mass). The depletion 
is stronger the lower the mass. As there seems to be no Be depletion for stars at ages of 50 and 150 Myr, we conclude that the 
Be depletion in this temperature range is taking place during the main sequence, in agreement with the conclusions of 
\citet{San04b} and \citet{Ran07}.

%

\subsection{Comparison with the models}\label{sec:models}

\begin{figure}
\begin{centering}
\includegraphics[width=7cm]{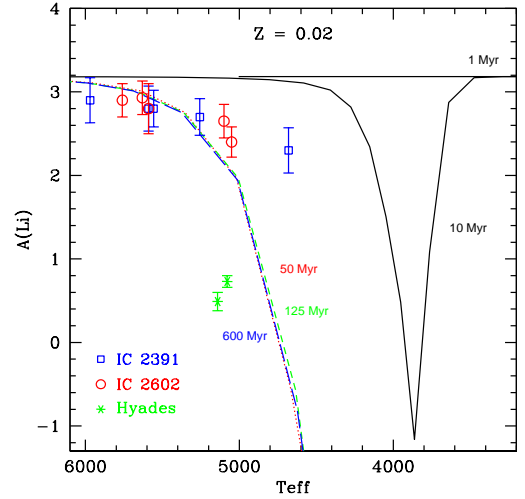}
\caption{Comparison of the Li abundances with the theoretical models of \citet{Siess00}. The curves show the Li abundance 
for stars with solar metallicity at the main sequence as a function of effective temperatures. Five different ages are shown: 1 and 10 Myr as black 
solid curves, 50 Myr as a red dotted curve, 125 Myr as a green short-dashed curve, and 600 Myr as a blue long-dashed curve. In 
addition to the stars of IC 2391, blue open squares, and IC 2602, red open circles (IC 2602), we show two stars from the Hyades as 
green starred symbols.}
\label{fig:limodel}
\end{centering}
\end{figure}

We now compare our abundances with expectations from theoretical models. Isochrones for the early-main sequence were 
calculated using the internet server of \citet{Siess00} for solar metallicity (Z = 0.02), with and without overshooting, 
and for 5 different ages (1, 10, 50, 125, and 600 Myr). In addition to the usual quantities (luminosity and 
temperature) the models offer additional information such as the surface abundance of 
the light elements (LiBeB).

When looking at the plots comparing abundances and models, it is important to remember that a star does not keep its 
temperature constant during the whole age interval for which we show models. In other words, a fixed temperature value 
in the following plots correspond to stars with different masses depending on the age. For example, at 1 Myr, a star with 
$\sim$ 3 $M_{\odot}$ has 5000 K, but at 50 Myr (onwards) a star with $\sim$ 0.90 $M_{\odot}$ has this temperature.

\begin{figure}
\begin{centering}
\includegraphics[width=7cm]{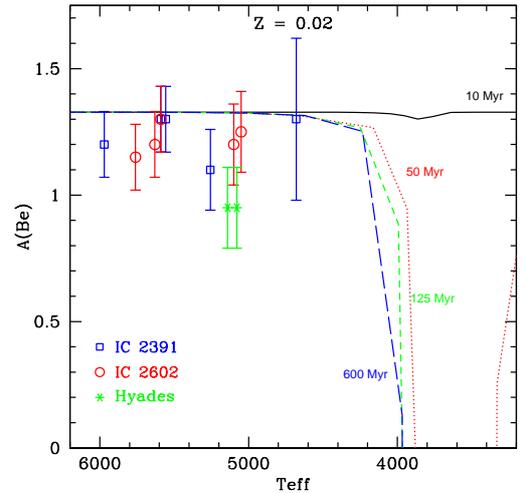}
\caption{Comparison of the Be abundances with the theoretical models of \citet{Siess00}. The curves show the Be abundance 
for stars with solar metallicity at the main sequence as a function of effective temperatures. Four different ages are shown: 10 Myr as a black 
solid curve, 50 Myr as a red dotted curve, 125 Myr as a green short-dashed curve, and 600 Myr as a blue long-dashed curve. 
Symbols for the stars are as in Fig. \ref{fig:limodel}.}
\label{fig:bemodel}
\end{centering}
\end{figure}

In Fig. \ref{fig:limodel} we compare the Li abundances with the models without overshooting. Both IC 2391 and IC 2602 have 
ages of about 50 Myr while the Hyades has $\sim$ 600 Myr. The plot shows what was discussed before, i. e., the models 
predict too much Li depletion for cool stars and no depletion during the main sequence. The difference between the 
young clusters and the Hyades, however, suggest that some depletion took place during the main sequence in this older cluster.

The same comparison is done with the Be abundances in Fig. \ref{fig:bemodel}. The models show slightly higher Be abundance 
because of a difference in the choice of reference solar abundance. Apart from that, the observations agree very well with the 
models. In the temperature range investigated, which for an age of 50 Myr correspond to a mass range between 
0.80 $\leq$ $M/M_{\odot}$ $\leq$ 1.20, no Be depletion is expected and none is seen.

Formally, the inclusion of the Hyades stars in these plots and their comparison with the Z = 0.02 models could be questioned
 as they are more metal rich ([Fe/H] = +0.13). From the models point of view, metallicity is expected to influence the degree of light 
 elements depletion \citep[e.g.][]{PiauTurck02,Sestito06}. An increased opacity would affect the conditions at the bottom of the convective zone during 
 PMS. In \citet{Ran07}, however, models of PMS Be depletion were calculated for both the solar and the Hyades composition for a 
star of 0.9 $M_{\odot}$. It was shown that the effect caused by metallicity in this case is negligible. Thus, the observed difference 
between the Be abundances can not be explained by the difference in metallicity. \citet{Sestito06} has indeed investigated in detail 
the effect of heavy-element opacity on PMS Li depletion. They concluded that changes in the abundances are not enough to 
result in agreement between models and Li observations in low-mass stars of young clusters. Observationally, no significant 
difference between Li depletion patterns has been found among clusters with similar ages and different metallicities, arguing 
against a strong metallicity dependence of PMS depletion \citep{JeffriesJames99,Jeffries06}.

\begin{figure}
\begin{centering}
\includegraphics[width=7cm]{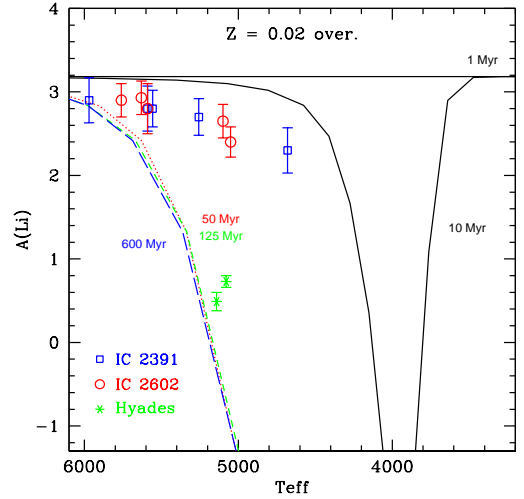}
\caption{Comparison of the Li abundances with the theoretical models of \citet{Siess00} including overshooting. Lines and 
symbols are as in Fig. \ref{fig:limodel}.}
\label{fig:liover}
\end{centering}
\end{figure}

\begin{figure}
\begin{centering}
\includegraphics[width=7cm]{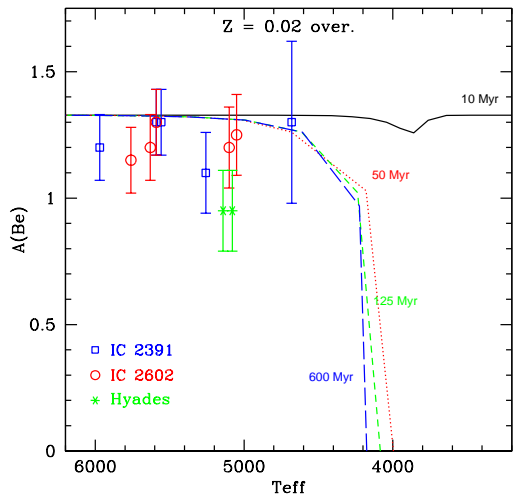}
\caption{Comparison of the Be abundances with the theoretical models of \citet{Siess00} including overshooting. Lines and 
symbols are as in Fig. \ref{fig:bemodel}.}
\label{fig:beover}
\end{centering}
\end{figure}

Similar comparisons are shown in Figs. \ref{fig:liover} and \ref{fig:beover} but against the models including overshooting. The 
discrepancy with the Li data is shown to increase when overshooting is taken into account. For Be however, the inclusion of 
overshooting is still not enough to cause Be depletion in the range of parameters of our sample. From these plots we see 
that we were still not able to reach down to effective temperatures where Be depletion during the pre-main sequence 
phases could be detected. If one can push the determination of Be abundances to even lower temperatures, 
it would be possible to look for the effect of overshooting on Be depletion. Such measurement are, however, be very challenging.

As discussed previously in \citet{Ran07}, the small difference seen between the Be abundances in the Hyades and in 
IC 2391/2602, seems to point to yet another shortcoming of the models. The mixing mechanism depleting Li during the main 
sequence seems also to be able to deplete Be, although not as dramatically.


\subsection{Evolution of Be in the disk}

The average Be abundance in the two clusters is A(Be) = 1.24 $\pm$ 0.09 and 1.22 $\pm$ 0.06, for IC 2391 and IC2602 respectively.
 These values can be compared to the average of other young clusters to search for variations of the Be abundance in the disk 
 and to investigate signs of Be evolution with time in the Galaxy. For this comparison we selected from the literature only stars with 
 5400 K $\leq$ $T_\mathrm{eff}$ $\leq$ 5800 K, i.e. stars that show no signs of Be depletion. For the clusters analyzed by Boesgaard and collaborators
  the averages are: A(Be) = 1.33 $\pm$ 0.06 for 10 stars of the 600 Myr old Hyades; A(Be) = 
1.26 $\pm$ 0.07 for 3 stars of the 120 Myr old Pleiades; A(Be) = 1.28 for 2 stars of the 500 Myr old Coma; A(Be) = 1.25 for 1 stars of 
the 300 Myr old Ursa Major moving group \citep{BK02,BAK03,BAK03b,BAK04}. 
All these average values seem to be in agreement with each other, given the uncertainties. Although one 
should keep in mind that there is a small spread in metallicity between the clusters (from [Fe/H] = $-$0.09 for Coma and [Fe/H] = +0.13 
for the Hyades) and that the results are not exactly on the same scale as ours, when seen together they indicate a lack of strong variations in the initial Be abundance 
among the clusters. 


By comparing the Be abundance in these young clusters with the solar abundance, one can search for evidence of Be enrichment in 
the disk. This task if made difficult by a possible disagreement between the solar photospheric and meteoritic abundances, A(Be) = 1.10 and 
A(Be) = 1.42 respectively. This meteoritic value recommended by \citet{AG89} seems to indicate a photospheric depletion by a factor of two. 

\citet{BaBe98} and \citet{BBB01}, however, argued that by accounting for a possible missing near-UV continuum opacity the photospheric Be abundance 
increases to A(Be) = 1.41, and thus is brought into agreement with the meteoritic one. This conclusion is corroborated by a re-analysis 
of the Be photospheric solar abundance using 3D hydrodynamical models \citep{Asp04}, where a recommended value of A(Be) = 
1.38 is found. As discussed in \citet{Asp05} and \citet{Asplund09}, Be abundances in the Sun seem to be largely insensitive to NLTE 
and 3D effects and this increase is only due to the missing opacity. It is interesting however to notice that two recent re-determinations 
of the Be abundance in a CI chondrite meteorite yields 
a value of A(Be) = 1.32 \citep[][see also \citeauthor{ipLodders10} \citeyear{ipLodders10}]{MakishimaNakamura06}, which is smaller than 
the currently recommended photospheric value. The implications of this smaller value have not been discussed in the literature yet, but this lower value 
might suggest that the  `missing opacity' has been overestimated in the above analyses. In their analysis of Be abundances in 
solar analogues, \citet{Takeda11} reached similar conclusion.

%
%

%
%

As discussed before, the Be abundances we derived in the young clusters show no evidence of PMS depletion and 
are in agreement with the one we derived for the Sun. Thus we can draw the two following conclusions: (1) there is no evidence for significant Be 
depletion in the Sun, in agreement with the conclusions of \citet{Ran02,Ran07}; (2) the similarity of the abundances 
also indicate that there has been no significant Galactic enrichment of Be in the last $\sim$ 4.5 Gyr since the Sun 
was formed.


%

\section{Summary}\label{sec:final}

Beryllium abundances were calculated for ten confirmed members of the young pre-main sequence 
clusters IC 2391 and IC 2602. We also recalculated Be abundances for two main sequence stars of 
the Hyades. With a careful abundance analysis, we were able to provide actual detections 
for fast-rotating stars and for stars with effective temperatures down to $\sim$ 4700 K. These young stars ($\sim$ 50 Myr)
have just arrived in the main sequence. Therefore, any alteration in the surface abundance of the light elements 
Li and Be should have taken place during the PMS phase.

We show that all the stars analyzed have, within the uncertainties, the same Be abundance even though they 
have different degrees of Li depletion. This result confirms empirically for the first time the theoretical prediction that Be abundances are 
not affected by mixing during the PMS for stars in the range between 0.80 $\leq$ $M/M_{\odot}$ $\leq$ 1.20, as 
expected by the models. In addition, as discussed in the literature, for stars less massive than the Sun the mixing mechanism depleting Li during the main 
sequence seems also to be able to deplete Be.

At even lower temperatures ($\sim$ 4000 K), models with and without overshooting differ in the predicted amount of Be depletion. 
It is not clear, however, if the Be abundances can be properly determined in such cool stars. The attempt to push the determination of 
Be abundances to even lower temperatures, although very challenging, could provide important information to differentiate 
between these models, testing the importance of overshooting during the PMS.  

\begin{acknowledgements}
This research has made use of the WEBDA database, operated at the Institute for Astronomy of the University of Vienna, of NASA's 
Astrophysics Data System, and of the Simbad database operated at CDS, Strasbourg, France.
\end{acknowledgements}

\bibliographystyle{aa}
\bibliography{../../Tese/rsmiljanic}

\begin{thebibliography}{93}
\expandafter\ifx\csname natexlab\endcsname\relax\def\natexlab#1{#1}\fi

\bibitem[{{Anders} \& {Grevesse}(1989)}]{AG89}
{Anders}, E. \& {Grevesse}, N. 1989, \gca, 53, 197

\bibitem[{{Asplund}(2004)}]{Asp04}
{Asplund}, M. 2004, \aap, 417, 769

\bibitem[{{Asplund}(2005)}]{Asp05}
{Asplund}, M. 2005, \araa, 43, 481

\bibitem[{Asplund {et~al.}(2009)Asplund, Grevesse, Sauval, \&
  Scott}]{Asplund09}
Asplund, M., Grevesse, N., Sauval, A.~J., \& Scott, P. 2009, \araa, 47, 481

\bibitem[{{Balachandran} {et~al.}(1988){Balachandran}, {Lambert}, \&
  {Stauffer}}]{Balachandran88}
{Balachandran}, S., {Lambert}, D.~L., \& {Stauffer}, J.~R. 1988, \apj, 333, 267

\bibitem[{{Balachandran} \& {Bell}(1998)}]{BaBe98}
{Balachandran}, S.~C. \& {Bell}, R.~A. 1998, Nature, 392, 791

\bibitem[{{Balachandran} {et~al.}(2011){Balachandran}, {Mallik}, \&
  {Lambert}}]{Balachandran11}
{Balachandran}, S.~C., {Mallik}, S.~V., \& {Lambert}, D.~L. 2011, \mnras, 410,
  2526

\bibitem[{{Ballester} {et~al.}(2000){Ballester}, {Modigliani}, {Boitquin},
  {Cristiani}, {Hanuschik}, {Kaufer}, \& {Wolf}}]{uvespipeline}
{Ballester}, P., {Modigliani}, A., {Boitquin}, O., {et~al.} 2000, The
  Messenger, 101, 31

\bibitem[{{Baraffe} \& {Chabrier}(2010)}]{BaraffeChabrier10}
{Baraffe}, I. \& {Chabrier}, G. 2010, \aap, 521, A44+

\bibitem[{{Barrado y Navascu{\'e}s} {et~al.}(2001{\natexlab{a}}){Barrado y
  Navascu{\'e}s}, {Deliyannis}, \& {Stauffer}}]{Barrado01}
{Barrado y Navascu{\'e}s}, D., {Deliyannis}, C.~P., \& {Stauffer}, J.~R.
  2001{\natexlab{a}}, \apj, 549, 452

\bibitem[{{Barrado y Navascu{\'e}s} {et~al.}(2001{\natexlab{b}}){Barrado y
  Navascu{\'e}s}, {Garc{\'{\i}}a L{\'o}pez}, {Severino}, \&
  {Gomez}}]{Barrado01b}
{Barrado y Navascu{\'e}s}, D., {Garc{\'{\i}}a L{\'o}pez}, R.~J., {Severino},
  G., \& {Gomez}, M.~T. 2001{\natexlab{b}}, \aap, 371, 652

\bibitem[{{Barrado y Navascu{\'e}s} {et~al.}(2004){Barrado y Navascu{\'e}s},
  {Stauffer}, \& {Jayawardhana}}]{Barrado04}
{Barrado y Navascu{\'e}s}, D., {Stauffer}, J.~R., \& {Jayawardhana}, R. 2004,
  \apj, 614, 386

\bibitem[{{Bell} {et~al.}(2001){Bell}, {Balachandran}, \& {Bautista}}]{BBB01}
{Bell}, R.~A., {Balachandran}, S.~C., \& {Bautista}, M. 2001, \apjl, 546, L65

\bibitem[{{Bildsten} {et~al.}(1997){Bildsten}, {Brown}, {Matzner}, \&
  {Ushomirsky}}]{Bildsten97}
{Bildsten}, L., {Brown}, E.~F., {Matzner}, C.~D., \& {Ushomirsky}, G. 1997,
  \apj, 482, 442

\bibitem[{{Bodenheimer}(1966)}]{Bodenheimer66}
{Bodenheimer}, P. 1966, \apj, 144, 103

\bibitem[{{Boesgaard} {et~al.}(2003{\natexlab{a}}){Boesgaard}, {Armengaud}, \&
  {King}}]{BAK03}
{Boesgaard}, A.~M., {Armengaud}, E., \& {King}, J.~R. 2003{\natexlab{a}}, \apj,
  583, 955

\bibitem[{{Boesgaard} {et~al.}(2003{\natexlab{b}}){Boesgaard}, {Armengaud}, \&
  {King}}]{BAK03b}
{Boesgaard}, A.~M., {Armengaud}, E., \& {King}, J.~R. 2003{\natexlab{b}}, \apj,
  582, 410

\bibitem[{{Boesgaard} {et~al.}(2004){Boesgaard}, {Armengaud}, \&
  {King}}]{BAK04}
{Boesgaard}, A.~M., {Armengaud}, E., \& {King}, J.~R. 2004, \apj, 605, 864

\bibitem[{{Boesgaard} \& {King}(2002)}]{BK02}
{Boesgaard}, A.~M. \& {King}, J.~R. 2002, \apj, 565, 587

\bibitem[{{Bouvier}(2008)}]{Bouvier08}
{Bouvier}, J. 2008, \aap, 489, L53

\bibitem[{{Butler} {et~al.}(1987){Butler}, {Marcy}, {Cohen}, \&
  {Duncan}}]{Butler87}
{Butler}, R.~P., {Marcy}, G.~W., {Cohen}, R.~D., \& {Duncan}, D.~K. 1987,
  \apjl, 319, L19

\bibitem[{{Castelli} \& {Kurucz}(2003)}]{ipCK03}
{Castelli}, F. \& {Kurucz}, R.~L. 2003, in Proceedings of the IAU Symposium
  210: Modelling of Stellar Atmospheres, ed. N.~{Piskunov}, W.~W. {Weiss}, \&
  D.~F. {Gray} (ASP, San Francisco), A20

\bibitem[{{Castilho} {et~al.}(1999){Castilho}, {Spite}, {Barbuy}, {Spite}, {De
  Medeiros}, \& {Gregorio-Hetem}}]{Cast99}
{Castilho}, B.~V., {Spite}, F., {Barbuy}, B., {et~al.} 1999, \aap, 345, 249

\bibitem[{{Chaboyer} {et~al.}(1995){Chaboyer}, {Demarque}, \&
  {Pinsonneault}}]{Chaboyer95}
{Chaboyer}, B., {Demarque}, P., \& {Pinsonneault}, M.~H. 1995, \apj, 441, 876

\bibitem[{{Chabrier} \& {Baraffe}(1997)}]{ChabrierBaraffe97}
{Chabrier}, G. \& {Baraffe}, I. 1997, \aap, 327, 1039

\bibitem[{{Chabrier} {et~al.}(2007){Chabrier}, {Gallardo}, \&
  {Baraffe}}]{Chabrier07}
{Chabrier}, G., {Gallardo}, J., \& {Baraffe}, I. 2007, \aap, 472, L17

\bibitem[{{Charbonnel} {et~al.}(2000){Charbonnel}, {Deliyannis}, \&
  {Pinsonneault}}]{ipCDP00}
{Charbonnel}, C., {Deliyannis}, C.~P., \& {Pinsonneault}, M. 2000, in IAU
  Symposium 198: The Light Elements and their Evolution, ed. L.~{da Silva},
  R.~{De Medeiros}, \& M.~{Spite} (ASP, San Francisco), 87--+

\bibitem[{{Chmielewski} {et~al.}(1975){Chmielewski}, {Brault}, \&
  {Mueller}}]{Chm75}
{Chmielewski}, Y., {Brault}, J.~W., \& {Mueller}, E.~A. 1975, \aap, 42, 37

\bibitem[{{Coelho} {et~al.}(2005){Coelho}, {Barbuy}, {Mel\'endez}, {Schiavon},
  \& {Castilho}}]{CB05}
{Coelho}, P., {Barbuy}, B., {Mel\'endez}, J., {Schiavon}, R.~P., \& {Castilho},
  B.~V. 2005, \aap, 443, 735

\bibitem[{{Corliss} \& {Bozman}(1962)}]{CorlissBozman62}
{Corliss}, C.~H. \& {Bozman}, W.~R. 1962, {Experimental transition
  probabilities for spectral lines of seventy elements} (US Dep. of Commerce,
  National Bureau of Standards, Washington)

\bibitem[{{da Silva} {et~al.}(2009){da Silva}, {Torres}, {de La Reza}, {Quast},
  {Melo}, \& {Sterzik}}]{DaSilva09}
{da Silva}, L., {Torres}, C.~A.~O., {de La Reza}, R., {et~al.} 2009, \aap, 508,
  833

\bibitem[{{Dekker} {et~al.}(2000){Dekker}, {D'Odorico}, {Kaufer}, {Delabre}, \&
  {Kotzlowski}}]{ipDe00}
{Dekker}, H., {D'Odorico}, S., {Kaufer}, A., {Delabre}, B., \& {Kotzlowski}, H.
  2000, in Proc. SPIE, Optical and IR Telescope Instrumentation and Detectors,
  ed. M.~{Iye} \& A.~F. {Moorwood}, Vol. 4008 (SPIE), 534--545

\bibitem[{{Deliyannis} {et~al.}(2000){Deliyannis}, {Pinsonneault}, \&
  {Charbonnel}}]{ipDPC00}
{Deliyannis}, C.~P., {Pinsonneault}, M.~H., \& {Charbonnel}, C. 2000, in IAU
  Symposium 198: The Light Elements and their Evolution, ed. L.~{da Silva},
  R.~{De Medeiros}, \& M.~{Spite} (ASP, San Francisco), 61

\bibitem[{{Dobbie} {et~al.}(2010){Dobbie}, {Lodieu}, \& {Sharp}}]{Dobbie10}
{Dobbie}, P.~B., {Lodieu}, N., \& {Sharp}, R.~G. 2010, \mnras, 409, 1002

\bibitem[{{D'Orazi} \& {Randich}(2009)}]{DoraziRandich09}
{D'Orazi}, V. \& {Randich}, S. 2009, \aap, 501, 553

\bibitem[{{Duncan} \& {Jones}(1983)}]{DuncanJones83}
{Duncan}, D.~K. \& {Jones}, B.~F. 1983, \apj, 271, 663

\bibitem[{{Ford} {et~al.}(2002){Ford}, {Jeffries}, \& {Smalley}}]{Ford02}
{Ford}, A., {Jeffries}, R.~D., \& {Smalley}, B. 2002, \aap, 391, 253

\bibitem[{{Garcia Lopez} {et~al.}(1994){Garcia Lopez}, {Rebolo}, \&
  {Martin}}]{GarciaLopez94}
{Garcia Lopez}, R.~J., {Rebolo}, R., \& {Martin}, E.~L. 1994, \aap, 282, 518

\bibitem[{{Garcia Lopez} {et~al.}(1995){Garcia Lopez}, {Rebolo}, \& {Perez de
  Taoro}}]{GL95}
{Garcia Lopez}, R.~J., {Rebolo}, R., \& {Perez de Taoro}, M.~R. 1995, \aap,
  302, 184

\bibitem[{{H{\"u}nsch} {et~al.}(2004){H{\"u}nsch}, {Randich}, {Hempel},
  {Weidner}, \& {Schmitt}}]{Hunsch04}
{H{\"u}nsch}, M., {Randich}, S., {Hempel}, M., {Weidner}, C., \& {Schmitt},
  J.~H.~M.~M. 2004, \aap, 418, 539

\bibitem[{{James} \& {Jeffries}(1997)}]{JamesJeffries97}
{James}, D.~J. \& {Jeffries}, R.~D. 1997, \mnras, 292, 252

\bibitem[{{Jeffries}(1999)}]{Jeffries99}
{Jeffries}, R.~D. 1999, \mnras, 309, 189

\bibitem[{{Jeffries}(2006)}]{Jeffries06}
{Jeffries}, R.~D. 2006, in Chemical Abundances and Mixing in Stars in the Milky
  Way and its Satellites, ESO Astrophysics Symposia, ed. {Randich, S.~\&
  Pasquini, L.} (Springer-Verlag, Berlin-Heidelberg), 163

\bibitem[{{Jeffries} {et~al.}(2009){Jeffries}, {Jackson}, {James}, \&
  {Cargile}}]{Jeffries09}
{Jeffries}, R.~D., {Jackson}, R.~J., {James}, D.~J., \& {Cargile}, P.~A. 2009,
  \mnras, 400, 317

\bibitem[{{Jeffries} \& {James}(1999)}]{JeffriesJames99}
{Jeffries}, R.~D. \& {James}, D.~J. 1999, \apj, 511, 218

\bibitem[{{Jones} {et~al.}(1997){Jones}, {Fischer}, {Shetrone}, \&
  {Soderblom}}]{Jones97}
{Jones}, B.~F., {Fischer}, D., {Shetrone}, M., \& {Soderblom}, D.~R. 1997, \aj,
  114, 352

\bibitem[{{Jones} {et~al.}(1996){Jones}, {Shetrone}, {Fischer}, \&
  {Soderblom}}]{Jones96}
{Jones}, B.~F., {Shetrone}, M., {Fischer}, D., \& {Soderblom}, D.~R. 1996, \aj,
  112, 186

\bibitem[{{King} {et~al.}(1997){King}, {Deliyannis}, \& {Boesgaard}}]{KDB97}
{King}, J.~R., {Deliyannis}, C.~P., \& {Boesgaard}, A.~M. 1997, \apj, 478, 778

\bibitem[{{King} {et~al.}(2000){King}, {Krishnamurthi}, \&
  {Pinsonneault}}]{King00}
{King}, J.~R., {Krishnamurthi}, A., \& {Pinsonneault}, M.~H. 2000, \aj, 119,
  859

\bibitem[{{King} {et~al.}(2010){King}, {Schuler}, {Hobbs}, \&
  {Pinsonneault}}]{King10}
{King}, J.~R., {Schuler}, S.~C., {Hobbs}, L.~M., \& {Pinsonneault}, M.~H. 2010,
  \apj, 710, 1610

\bibitem[{{Kurucz}(1993)}]{Kuruczcd13}
{Kurucz}, R. 1993, ATLAS9 Stellar Atmosphere Programs and 2 km/s grid. CD-ROM
  No.~13.~ Cambridge, Mass.: Smithsonian Astrophysical Observatory.

\bibitem[{{Kurucz}(1988)}]{inpKurucz88}
{Kurucz}, R.~L. 1988, in Transactions of the International Astronomical Union,
  Volume XXB, ed. M.~{McNally} (Kluwer, Dordrecht), 168--172

\bibitem[{{Lodders}(2010)}]{ipLodders10}
{Lodders}, K. 2010, in Principles and Perspectives in Cosmochemistry,
  Astrophysics and Space Science, ed. {A.~Goswami \& B.~E.~Reddy} (Springer,
  Berlin), 379--+

\bibitem[{{MacDonald} \& {Mullan}(2010)}]{MacdonaldMullan10}
{MacDonald}, J. \& {Mullan}, D.~J. 2010, \apj, 723, 1599

\bibitem[{{Makishima} \& {Nakamura}(2006)}]{MakishimaNakamura06}
{Makishima}, A. \& {Nakamura}, E. 2006, Geostandards and Geoanalytical
  Research, 30, 245

\bibitem[{{Martin} \& {Claret}(1996)}]{MartinClaret96}
{Martin}, E.~L. \& {Claret}, A. 1996, \aap, 306, 408

\bibitem[{{Martin} \& {Montes}(1997)}]{MartinMontes97}
{Martin}, E.~L. \& {Montes}, D. 1997, \aap, 318, 805

\bibitem[{{Mendes} {et~al.}(1999){Mendes}, {D'Antona}, \&
  {Mazzitelli}}]{Mendes99}
{Mendes}, L.~T.~S., {D'Antona}, F., \& {Mazzitelli}, I. 1999, \aap, 341, 174

\bibitem[{{Mermilliod} {et~al.}(2009){Mermilliod}, {Mayor}, \&
  {Udry}}]{Mermilliod09}
{Mermilliod}, J.-C., {Mayor}, M., \& {Udry}, S. 2009, \aap, 498, 949

\bibitem[{{Patten} \& {Simon}(1996)}]{PattenSimon96}
{Patten}, B.~M. \& {Simon}, T. 1996, \apjs, 106, 489

\bibitem[{{Piau} \& {Turck-Chi{\`e}ze}(2002)}]{PiauTurck02}
{Piau}, L. \& {Turck-Chi{\`e}ze}, S. 2002, \apj, 566, 419

\bibitem[{{Pinsonneault}(1997)}]{Pin97}
{Pinsonneault}, M. 1997, \araa, 35, 557

\bibitem[{{Platais} {et~al.}(2007){Platais}, {Melo}, {Mermilliod},
  {Kozhurina-Platais}, {Fulbright}, {M{\'e}ndez}, {Altmann}, \&
  {Sperauskas}}]{Platais07}
{Platais}, I., {Melo}, C., {Mermilliod}, J., {et~al.} 2007, \aap, 461, 509

\bibitem[{{Primas} {et~al.}(1997){Primas}, {Duncan}, {Pinsonneault},
  {Deliyannis}, \& {Thorburn}}]{Pr97}
{Primas}, F., {Duncan}, D.~K., {Pinsonneault}, M.~H., {Deliyannis}, C.~P., \&
  {Thorburn}, J.~A. 1997, \apj, 480, 784

\bibitem[{{Prosser} {et~al.}(1996){Prosser}, {Randich}, \&
  {Stauffer}}]{Prosser96}
{Prosser}, C.~F., {Randich}, S., \& {Stauffer}, J.~R. 1996, \aj, 112, 649

\bibitem[{{Randich}(2001)}]{Randich01}
{Randich}, S. 2001, \aap, 377, 512

\bibitem[{{Randich} {et~al.}(1997){Randich}, {Aharpour}, {Pallavicini},
  {Prosser}, \& {Stauffer}}]{Randich97}
{Randich}, S., {Aharpour}, N., {Pallavicini}, R., {Prosser}, C.~F., \&
  {Stauffer}, J.~R. 1997, \aap, 323, 86

\bibitem[{{Randich} {et~al.}(1998){Randich}, {Martin}, {Lopez}, \&
  {Pallavicini}}]{RMLP98}
{Randich}, S., {Martin}, E.~L., {Lopez}, R.~J.~G., \& {Pallavicini}, R. 1998,
  \aap, 333, 591

\bibitem[{{Randich} {et~al.}(2001){Randich}, {Pallavicini}, {Meola},
  {Stauffer}, \& {Balachandran}}]{Ran01}
{Randich}, S., {Pallavicini}, R., {Meola}, G., {Stauffer}, J.~R., \&
  {Balachandran}, S.~C. 2001, \aap, 372, 862

\bibitem[{{Randich} {et~al.}(2002){Randich}, {Primas}, {Pasquini}, \&
  {Pallavicini}}]{Ran02}
{Randich}, S., {Primas}, F., {Pasquini}, L., \& {Pallavicini}, R. 2002, \aap,
  387, 222

\bibitem[{{Randich} {et~al.}(2007){Randich}, {Primas}, {Pasquini}, {Sestito},
  \& {Pallavicini}}]{Ran07}
{Randich}, S., {Primas}, F., {Pasquini}, L., {Sestito}, P., \& {Pallavicini},
  R. 2007, \aap, 469, 163

\bibitem[{{Randich} {et~al.}(1995){Randich}, {Schmitt}, {Prosser}, \&
  {Stauffer}}]{Randich95}
{Randich}, S., {Schmitt}, J.~H.~M.~M., {Prosser}, C.~F., \& {Stauffer}, J.~R.
  1995, \aap, 300, 134

\bibitem[{{Santos} {et~al.}(2004){Santos}, {Israelian}, {Randich},
  {Garc{\'{\i}}a L{\'o}pez}, \& {Rebolo}}]{San04b}
{Santos}, N.~C., {Israelian}, G., {Randich}, S., {Garc{\'{\i}}a L{\'o}pez},
  R.~J., \& {Rebolo}, R. 2004, \aap, 425, 1013

\bibitem[{{Sbordone}(2005)}]{Sbordone05}
{Sbordone}, L. 2005, Memorie della Societa Astronomica Italiana Supplement, 8,
  61

\bibitem[{{Sbordone} {et~al.}(2004){Sbordone}, {Bonifacio}, {Castelli}, \&
  {Kurucz}}]{Sbordone04}
{Sbordone}, L., {Bonifacio}, P., {Castelli}, F., \& {Kurucz}, R.~L. 2004,
  Memorie della Societa Astronomica Italiana Supplement, 5, 93

\bibitem[{{Sestito} {et~al.}(2006){Sestito}, {Degl'Innocenti}, {Prada Moroni},
  \& {Randich}}]{Sestito06}
{Sestito}, P., {Degl'Innocenti}, S., {Prada Moroni}, P.~G., \& {Randich}, S.
  2006, \aap, 454, 311

\bibitem[{{Sestito} {et~al.}(2003){Sestito}, {Randich}, {Mermilliod}, \&
  {Pallavicini}}]{Sestito03}
{Sestito}, P., {Randich}, S., {Mermilliod}, J.-C., \& {Pallavicini}, R. 2003,
  \aap, 407, 289

\bibitem[{{Siess} {et~al.}(2000){Siess}, {Dufour}, \& {Forestini}}]{Siess00}
{Siess}, L., {Dufour}, E., \& {Forestini}, M. 2000, \aap, 358, 593

\bibitem[{{Smiljanic} {et~al.}(2009){Smiljanic}, {Pasquini}, {Bonifacio},
  {Galli}, {Gratton}, {Randich}, \& {Wolff}}]{Sm09}
{Smiljanic}, R., {Pasquini}, L., {Bonifacio}, P., {et~al.} 2009, \aap, 499, 103

\bibitem[{{Smiljanic} {et~al.}(2010){Smiljanic}, {Pasquini}, {Charbonnel}, \&
  {Lagarde}}]{Sm10}
{Smiljanic}, R., {Pasquini}, L., {Charbonnel}, C., \& {Lagarde}, N. 2010, \aap,
  510, A50

\bibitem[{{Smiljanic} {et~al.}(2008){Smiljanic}, {Pasquini}, {Primas},
  {Mazzali}, {Galli}, \& {Valle}}]{Sm08}
{Smiljanic}, R., {Pasquini}, L., {Primas}, F., {et~al.} 2008, \mnras, 385, L93

\bibitem[{{Soderblom} {et~al.}(1995){Soderblom}, {Jones}, {Stauffer}, \&
  {Chaboyer}}]{Soderblom95}
{Soderblom}, D.~R., {Jones}, B.~F., {Stauffer}, J.~R., \& {Chaboyer}, B. 1995,
  \aj, 110, 729

\bibitem[{{Soderblom} {et~al.}(1990){Soderblom}, {Oey}, {Johnson}, \&
  {Stone}}]{Soderblom90}
{Soderblom}, D.~R., {Oey}, M.~S., {Johnson}, D.~R.~H., \& {Stone}, R.~P.~S.
  1990, \aj, 99, 595

\bibitem[{{Soderblom} {et~al.}(1993){Soderblom}, {Pilachowski}, {Fedele}, \&
  {Jones}}]{Soderblom93b}
{Soderblom}, D.~R., {Pilachowski}, C.~A., {Fedele}, S.~B., \& {Jones}, B.~F.
  1993, \aj, 105, 2299

\bibitem[{{Stauffer} {et~al.}(1989){Stauffer}, {Hartmann}, {Jones}, \&
  {McNamara}}]{Stauffer89}
{Stauffer}, J., {Hartmann}, L.~W., {Jones}, B.~F., \& {McNamara}, B.~R. 1989,
  \apj, 342, 285

\bibitem[{{Stauffer} {et~al.}(1997){Stauffer}, {Hartmann}, {Prosser},
  {Randich}, {Balachandran}, {Patten}, {Simon}, \& {Giampapa}}]{Stauffer97}
{Stauffer}, J.~R., {Hartmann}, L.~W., {Prosser}, C.~F., {et~al.} 1997, \apj,
  479, 776

\bibitem[{{Stephens} {et~al.}(1997){Stephens}, {Boesgaard}, {King}, \&
  {Deliyannis}}]{SBKD97}
{Stephens}, A., {Boesgaard}, A.~M., {King}, J.~R., \& {Deliyannis}, C.~P. 1997,
  \apj, 491, 339

\bibitem[{{Stuik} {et~al.}(1997){Stuik}, {Bruls}, \& {Rutten}}]{Stuik97}
{Stuik}, R., {Bruls}, J.~H.~M.~J., \& {Rutten}, R.~J. 1997, \aap, 322, 911

\bibitem[{{St{\"u}tz} {et~al.}(2006){St{\"u}tz}, {Bagnulo}, {Jehin}, {Ledoux},
  {Cabanac}, {Melo}, \& {Smoker}}]{Stutz06}
{St{\"u}tz}, C., {Bagnulo}, S., {Jehin}, E., {et~al.} 2006, \aap, 451, 285

\bibitem[{{Takeda} {et~al.}(2011){Takeda}, {Tajitsu}, {Honda}, {Kawanomoto},
  {Ando}, \& {Sakurai}}]{Takeda11}
{Takeda}, Y., {Tajitsu}, A., {Honda}, S., {et~al.} 2011, \pasj, 63, 697

\bibitem[{{van Leeuwen}(2009)}]{vanLeeuwen09}
{van Leeuwen}, F. 2009, \aap, 497, 209

\bibitem[{{Ventura} {et~al.}(1998){Ventura}, {Zeppieri}, {Mazzitelli}, \&
  {D'Antona}}]{Ventura98}
{Ventura}, P., {Zeppieri}, A., {Mazzitelli}, I., \& {D'Antona}, F. 1998, \aap,
  331, 1011

\bibitem[{{Xiong} \& {Deng}(2005)}]{XiongDeng05}
{Xiong}, D. \& {Deng}, L. 2005, \apj, 622, 620

\end{thebibliography}

\end{document}